\documentclass[authoryear]{elsarticle}
\usepackage{graphicx} 
\usepackage{todonotes}
\usepackage{geometry}
\usepackage{float}
\usepackage{color}
\usepackage{soul}
\usepackage{amsmath}
\usepackage{subcaption}
\usepackage{amsmath,amssymb}
\usepackage[hidelinks]{hyperref}
\usepackage[nameinlink,capitalize,noabbrev]{cleveref} 
\usepackage{comment}
\usepackage{xcolor}

\title{Modeling Disruptions to Urban Metabolism \\using Interconnected Networks}

\author[1]{Bharat~Sharma}
\author[2]{Abhilasha~J.~Saroj}
\author[1]{Evan~Scherrer}
\author[1,*]{Melissa~R.~Allen-Dumas}

\affiliation[1]{organization = {Computational Sciences and Engineering Division,
Oak Ridge National Laboratory},
addressline = {Oak Ridge, TN 37830}, 
country={USA}}
\affiliation[2]{organization = {Buildings and Transportation Science Division,
Oak Ridge National Laboratory},
addressline = {Oak Ridge, TN 37830},
country={USA}}
\affiliation[*]{Corresponding author; allenmr@ornl.gov}

\begin{document}
\maketitle

\section*{Abstract} 
Representation of cities as organisms with metabolic processes is a useful analogy for urban design, development and sustainability. Urban metabolism can be modeled by representing urban systems as networks. The various networks included in a city’s metabolism are interdependent in complex ways. Thus, understanding the interaction among these networks is essential to understanding how a healthy urban metabolism is sustained and how injuries to the metabolic system can ``heal''. It is particularly important to understand how disruptions to one system in an urban area affect the functioning of other systems. Using distribution-level data from a real U.S. city on the electricity distribution system and road geometry, we apply connected network modeling to two critical inter-connected urban infrastructure sectors: energy and transportation. We quantify the robustness of these interdependent networks by evaluating the connectivity disruptions that may occur due to natural or synthetic disruptive events, using both unweighted and weighted metrics.


\vspace{.5cm}

\noindent\textbf{Keywords:} urban metabolism, network resilience, critical infrastructure, cascading failures, weighted networks, interdependency, power distribution, transportation networks

\section{Introduction and Background}\label{sec:introduction}
Many
\footnote{This manuscript has been authored by UT-Battelle LLC under contract DE-AC05-00OR22725 with the US Department of Energy (DOE). The US government retains and the publisher, by accepting the article for publication, acknowledges that the US government retains a nonexclusive, paid-up, irrevocable worldwide license to publish or reproduce the published form of this manuscript, or allow others to do so, for US government purposes. DOE will provide public access to these results of federally sponsored research in accordance with the DOE Public Access Plan (http://energy.gov/downloads/doe-public-access-plan).} 
researchers \citep[e.g.,][]{pincetl2009potential, kennedy2011study, burgess2015growth, wolman1965metabolism} have suggested that urban areas can be thought of as analogs to organisms with metabolic systems. Processes that make up a city’s metabolism are ecological, technical and socioeconomic, and the activity associated with these processes result in city growth, production of energy and elimination of waste \citep{kennedy2007changing}. \citet{ferrao2013sustainable} emphasize that each city's set of urban systems include traits that reflect their cultural and socioeconomic context as well as their interaction with local geography and natural surroundings \citep{ferrao2013sustainable}. Framing urban activity as metabolism allows for the explicit definition of city system boundaries, accounting for city input and output flows, and the hierarchical decomposition of independent and interdependent systems within city boundaries.

\citet{baccini1991anthroposphere} assessed cities’ input and output flows as mass fluxes of water, materials and nutrients. 
\citet{kennedy2011study} traced flows of water, energy, nutrients and materials through an urban system focusing on the reduction of input resources and output waste and the monitoring of pertinent information about energy efficiency, material cycling, and infrastructure in urban systems. \citet{soltan2015analysis} used a linearized graph-based power flow model to represent failures and cascades in the electrical grid to show how the disruption of the flow of urban energy caused consequences beyond an initial rupture in the system. The cascades they represented; that is, the initially localized events that further propagated \cite{hackett2011cascade}, were contained within the electrical grid system. Additionally,   \citet{haseltine2017prediction} developed a neural network tool to evaluate the performance of the components within an electrical system interconnection on a metric of overall reliability.

Urban size has been shown to be positively correlated with energy consumption \citep{chen2011estimating} because larger populations in cities contribute to increased energy demand in residential, commercial and industrial buildings, and in the travel modes by which large populations access these buildings. However, to some extent, large cities' population density concentrates resource consumption and waste emission so that better energy efficiency is achieved as compared to smaller cities  \citep{zhang2015urban}. However, more recent studies are working toward a ``generalized theory'' of urban metabolism making connections between city-scale systems and flow quantities to the covariation of the same characteristics within cities \citep{hendrick2025stochastic, pandey2025rising} across the city space.

Just as complex organisms contain many internal physical networks (e.g., blood vessels, nervous system, digestive tract), a city contains network properties that serve analogous functions (transportation, electric grid, waste disposal). For representing these flows as networks, the various components of the metabolic processes become nodes and flows travel on network ``edges'' between nodes represented within a network model using mathematical descriptions \citep{zhang2015urban}. Examples of single network models of the urban metabolic functions considered here include traffic network models, (e.g., \citep{samaniego2008cities, zhou2021intelligent, klaylee2023effect, rahimitouranposhti2025network}) and the bulk electric system \citep{lee2016urban, barrera2018multi}. Interconnected multi-network models have been developed by \cite{ganguly2020resilience, ganguly2023simulation, Warner2019} and others to study the urban systems.

\begin{figure}[H]
    \centering
    \includegraphics[height=0.35\textheight,keepaspectratio]{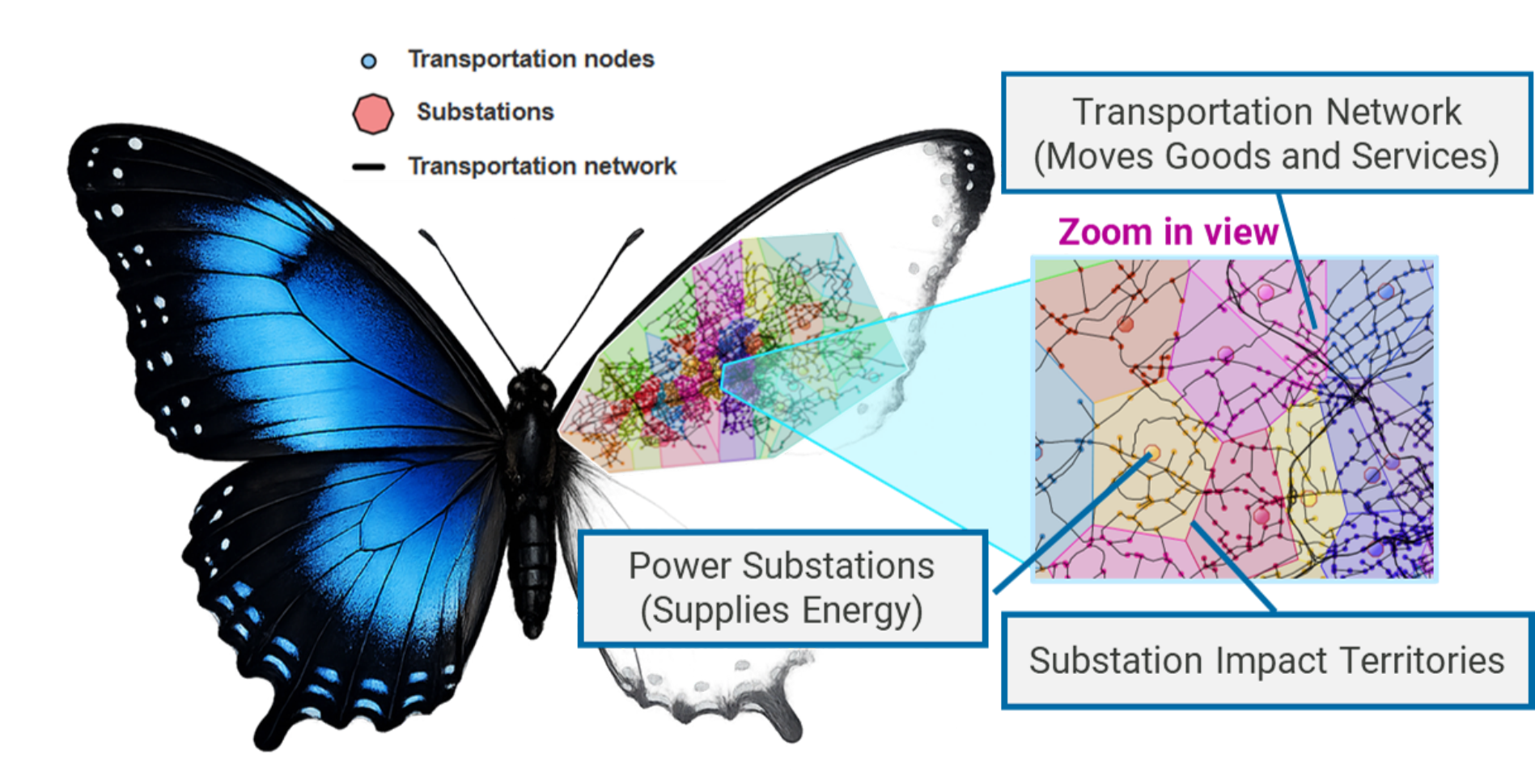}
    \vspace{-10pt}
    \caption{Conceptual illustration of urban interdependent electric grid and transportation as systems in an organism. The transportation network (bloodstream) that moves goods and services depends on traffic lights powered by electric substations (mitochondria) that supply energy. The boundaries show each substation’s impact territory over nearby transportation nodes.}
    \label{fig:butterfly_grid_transport_analogy}
\end{figure}

Some researchers divide urban metabolic processes into anabolic and catabolic \citep{costa2008general}. In this context, anabolic processes refer to the consumption of resources to produce products, while catabolic processes refer to the decomposition and recycling of wastes. Here, we focus on two anabolic components of city metabolism: energy and mobility, along with their independent and interdependent responses to environmental assault or ``injury'' and their ability to return to functionality (or resilience) once an assault occurs.
We approach this problem by characterizing the metabolic components as connected networks, with the mobility network represented by a transportation road network that may be affected by cascades from the city’s electric grid (energy) system to the transportation network. The analogy here is that the electric grid substations loosely represent cell mitochondria converting fuel to energy, and the road network approximates blood vessels delivering nutrients from the mitochondria to various locations in the body (See Figure \ref{fig:butterfly_grid_transport_analogy}). Road intersections become disrupted when electricity fails to power traffic lights, street lights, and communication equipment, disrupting the flow of goods (nutrients) throughout the city. Here we postulate that if a substation fails, the nearby area it serves loses power. The roads continue to exist, but intersections and other key points do not manage traffic as well. This condition leads to confusion at junctions, lower road capacity and longer delays than typical, and queues that propagate backward. This situation may be analogous to cell dysfunction as a result of mitochondrial failure, although blood vessels are still present.
By framing energy and mobility as interacting anabolic subsystems, this study treats infrastructure disruptions as metabolic injuries shaped by internal robustness and cross-system dependencies, motivating the following research questions:
\begin{enumerate}
\item How do weighted and unweighted robustness analyses differ in identifying critical infrastructure components, and what planning operational implications follow?

\item How do cascading failures propagate between interdependent electricity and transportation networks? 

\item What are the implications of these failures for the overall city as an organism's health? 
\end{enumerate}

\section{Experimental Design}\label{sec:experiment_design}
The experimental design for this paper builds on the previous work of \cite{scherrer2025analyzing}. This work is furthered by the addition of realistic traffic demand over the road network and the discussion of an impacts-to-overall-infrastructure-resiliency-and-functionality metric using a 
hypothetical city case study
The disruptions in the network are modeled as node removal scenarios. The case study shows how a simulated disruption to the electric grid impacts the flow of vehicles on the road network, limiting the transport of goods and services, including those supporting energy generation, in and out of the urban metabolic system.


\subsection{Data}\label{sec:data}
In this paper, we use hypothetical 1) transportation road geometry and traffic demand and 2) substation location data to investigate road network robustness to potential electric grid network disruptions. 

\subsubsection{Road Geometry and Traffic Demand}\label{sec:geometry_demand_data}
We generated a shapefile with 4,799 road segments and 3658 nodes and geometric attributes for the road network. 
The road network consists of collector roadways along with hypothetical primary roads, i.e., no local/residential streets. For these roads we estimated a representative traffic demand on each network edge (street segment between intersections also known as links; the intersections are treated as nodes the edges connect). 
To the network we allocated an average bidirectional flow of 9038 vehicles and a maximum flow of 104,253 vehicles between two nodes. This allocation allowed for a link-level traffic demand estimated with a standard \emph{user equilibrium} (UE) traffic assignment in TransCAD. In UE assignment, travelers are assumed to choose routes that minimize their own perceived travel cost, and the network reaches an equilibrium in which no traveler can unilaterally reduce individual travel time by switching routes. UE-based assignments are widely used in regional planning practice and travel-demand modeling to convert OD demand into link flows, and they form the foundation for many static traffic assignment methods implemented in commercial planning software \citep{sheffi1985,patriksson1994,florian1995}.

\subsubsection{Substation Locations}\label{sec:substation_locations}
The electric grid substation data 
generated for this study represent
high-voltage facilities within an electric grid system. Substations are connected by power lines to power generation stations, to other substations, and to end-use customers. Often they contain transformers that step voltage down from higher voltage lines to lower voltage lines. In some cases, substations convert alternating current to direct current or vice versa. For this study, substations are treated simply as unweighted nodes in the electric grid system and are placed at various locations more or less coincident with groups of road intersections.  

\subsection{Resilience Analysis Design}\label{sec: resilience_analysis_design}
For the resilience analysis, we first investigated the resilience of the traffic demand as both a weighted and an unweighted road transportation network independently using the centrality metrics and resilience metrics described in \cref{sec:centrality_metrics} and \cref{sec:reslience_metrics} in detail. Next, we connected the two networks - the road intersections and the substations to evaluate cascading network-on-network impacts.


\subsubsection{Connecting the Electric Substations and Transportation Nodes}\label{sec:interconnectivity}

\begin{figure}[H]
    \centering

    \begin{subfigure}[t]{0.95\textwidth}
        \centering
        \includegraphics[width=\textwidth, trim=0cm 0cm 0cm 1cm, clip]{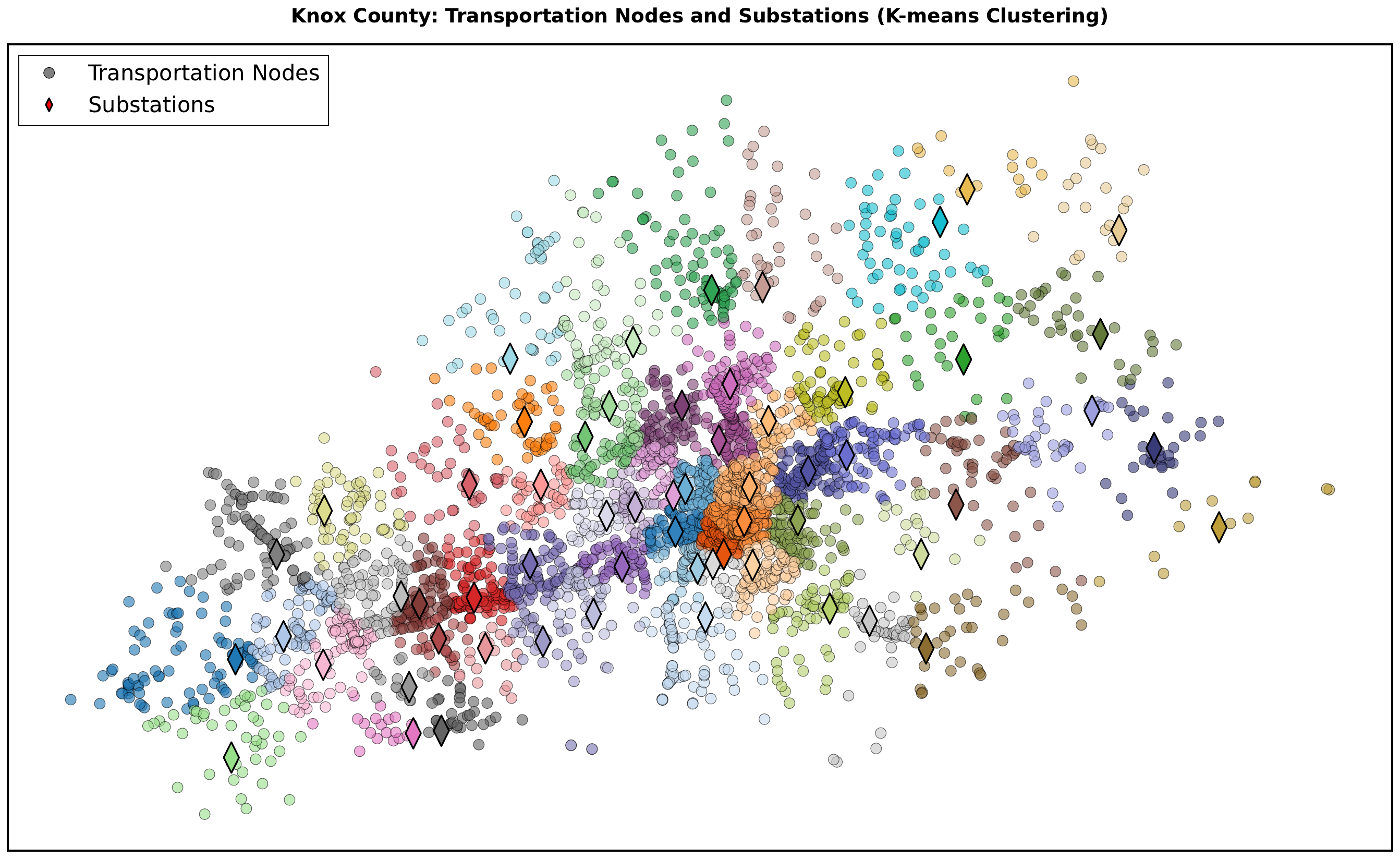}
        \caption{}
        \label{fig:BS_spatial_clusters_plot_a}
    \end{subfigure}

    \vspace{0.4cm}

    \begin{subfigure}[t]{0.95\textwidth}
        \centering
        \includegraphics[width=\textwidth, trim=0cm 0cm 0cm 1cm, clip]{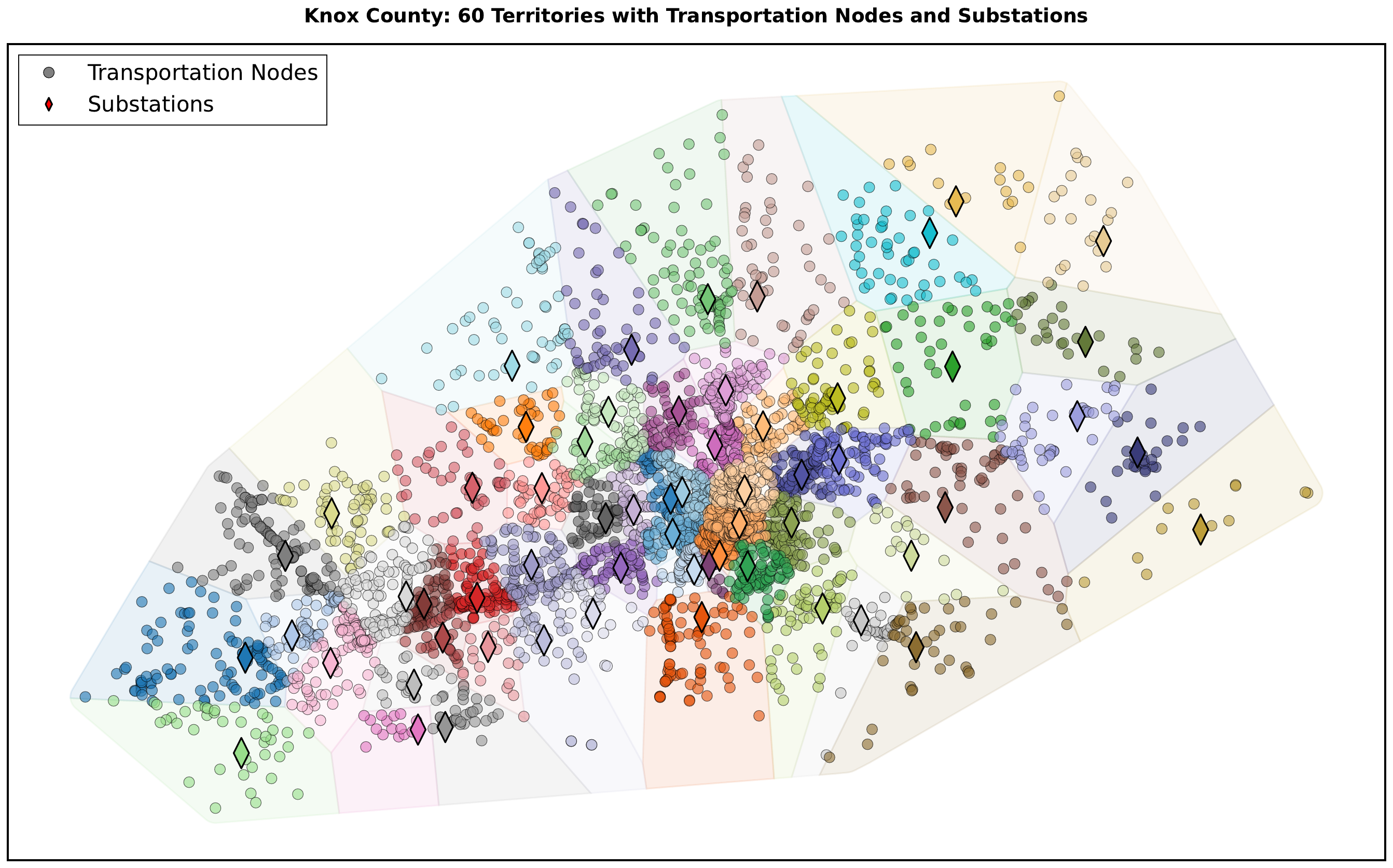}
        \caption{}
        \label{fig:60_vornoi_clustering_b}
    \end{subfigure}

    \caption{Cluster plots showing electric substations (diamonds) and their associated transportation nodes (circles) assigned to partitions derived using a) K-means clustering, and b) Voronoi tessellation}
    \label{fig:substation_transport_clustering_ab}
\end{figure}

We assigned geographic coordinates to all substations in the spatial bounds of the transportation network and to the geographic coordinates of the transportation nodes. To characterize spatial interdependence, we employed two clustering algorithms for allocating road network nodes to substation (as node) locations: 1) a specialized k-means clustering approach that deviates from the standard k-means implementation and 2) a Voronoi tessellation clustering approach. Figure~\ref{fig:substation_transport_clustering_ab} illustrates the resulting spatial clusters, showing transportation nodes color-coded according to their associated electric substation centroid. 

\emph{Specialized k-means clustering}:Rather than allowing cluster centroids to iterate and optimize through the standard k-means clustering algorithm, our methodology employed a constrained variant in which substation locations were designated as fixed centroids. This adaptation reflects the foundational principle that transportation infrastructure naturally organizes around electrical service points, rather than vice versa. The k-means clustering algorithm partitions observations into k clusters by minimizing the within-cluster sum of squared Euclidean distances between each data point and its nearest centroid. In standard k-means, both the assignment of points to clusters and the computation of centroid positions iterate until convergence. However, in our application, we fixed the cluster centers at substation locations from the outset, constraining the algorithm to perform only the assignment phase. Specifically, each transportation node was assigned to the nearest substation based on Euclidean distance in geographic space, creating spatial groups that reflect the principle of proximity-to-service-infrastructure. This methodological choice is justified by the hypothesis that transportation infrastructure and electrical infrastructure exhibit spatial interdependence, with transportation nodes tending to cluster around areas served by electrical substations.

\emph{Voronoi tessellation}:
A Voronoi tessellation partitions a domain into regions of influence around a set of \emph{sites}
(e.g. points representing facilities). We constructed service territories using a planar Voronoi tessellation, which partitions space into regions of influence around a set of point sites (here, substation/territory centroids) \citep{aurenhammer1991,okabe2000}. For sites $p_i\in\mathbb{R}^2$ and Euclidean distance $d(x,p_i)=\lVert x-p_i\rVert_2$, the Voronoi cell associated with site $p_i$ is
\begin{equation}
V_i=\left\{x\in\mathbb{R}^2 \mid \lVert x-p_i\rVert_2 \le \lVert x-p_j\rVert_2 \;\; \forall j\neq i\right\}.
\label{eq:voronoi_cell_points}
\end{equation}
\textcolor{black}{This yields a Voronoi tessellation whose cell boundaries are perpendicular bisectors between neighboring sites; under general position, the cells are convex polygons \citep{aurenhammer1991,okabe2000}.We used the centroid coordinates as generator points and constructed the Voronoi polygons using a standard Voronoi implementation (SciPy), consistent with classical constructions \citep{fortune1986,okabe2000}.} Since outer Voronoi regions may be unbounded, we added auxiliary points to create a large bounding box that induces finite polygons, and then restricted each territory to the study area by intersecting with a boundary polygon $B$ derived from the convex hull of the transportation network geometry. The resulting territories are, therefore, $\tilde{V}_i = V_i \cap B$. While this approach provides a useful geometric baseline for defining approximate influence areas \citep{okabe2000}, it assumes (i) Euclidean proximity governs assignment (rather than network travel time or electrical constraints), (ii) all sites have equal influence (i.e., no capacity/load weighting as in weighted Voronoi or power diagrams), and (iii) the convex-hull boundary and auxiliary points can affect territories near the domain edges \citep{aurenhammer1991,okabe2000}. For analyses requiring operational realism, these territories should be complemented with network-based distance measures or power-system planning models \citep{okabe2000}. Figure~\ref{fig:60_vornoi_clustering_b} visualizes the Voronoi service-territory boundaries as lightly shaded background regions. An interactive version of the visualization is at \url{https://abhilashasaroj17.github.io/Network-Analysis-Paper/}.

\section{Methods for Resilience Analysis}\label{sec:methods}
Robustness was assessed through sequential node removal experiments that simulate infrastructure failure and quantify resulting changes in network structure and performance. When removing nodes sequentially to simulate disruptions or failures, there are two fundamental approaches to centrality computation: a) initial (static) and b) recalculated (dynamic) centrality. In the former approach, centrality metrics are computed only once on the intact network or Giant Connected Component (GCC) at a time $t=0$ and all node removal decisions are based on this fixed ranking of centrality metric. Centrality metrics in the latter approach are recomputed after each node removal, reflecting the changing network topology.

\subsection{Weighted Transportation Network}
\label{sec:flow_weight}





\textcolor{black}{The AADT dataset provides bidirectional ``total\_flow'' for each link. We use this total flow to weight transportation edges, applying min--max normalization to map flows to $[0,1]$. Let $f_i$ be the total flow on link $i$; then}
\begin{equation}
\textcolor{black}{f_{\max}=\max_i(f_i),\qquad f_{\min}=\min_i(f_i)}
\end{equation}
\textcolor{black}{and the normalized flow weight is}
\begin{equation}
\textcolor{black}{fw_i=\frac{f_i-f_{\min}}{f_{\max}-f_{\min}}}
\label{eq:flow_weight}
\end{equation}
\textcolor{black}{This assigns $0$ to the minimum-flow link and $1$ to the maximum-flow link, with linear scaling in between (applied when $f_{\max}>0$).}

\begin{figure}[H]
    \centering
    \includegraphics[width=1\linewidth]{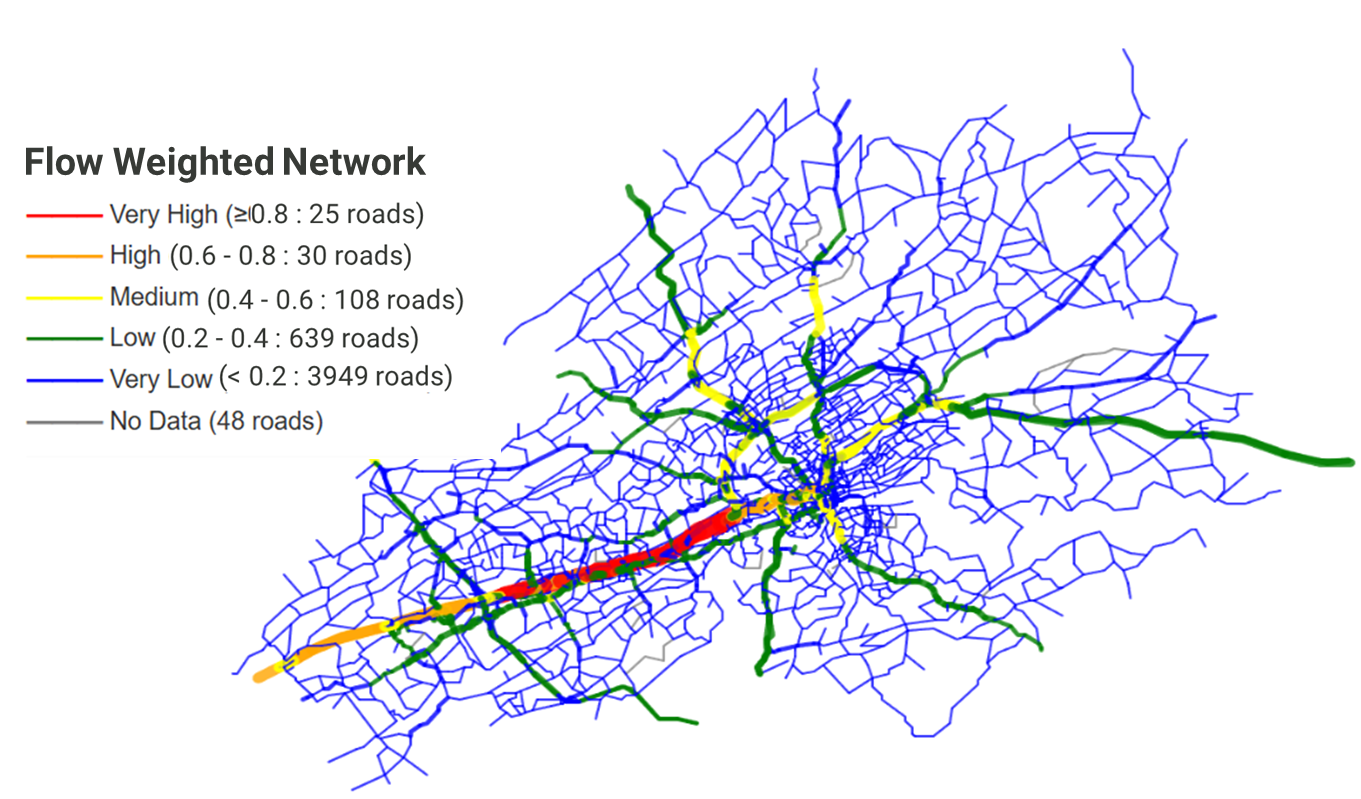}
    \caption{Flow weighted transportation network.}
    \label{fig:flow_weighted_transp}
\end{figure}

The flow-weighted network edges were then divided into five categories based on their weights: 
(1) Very High ($w \geq 0.8$), 
(2) High ($0.6 \leq w < 0.8$), 
(3) Medium ($0.4 \leq w < 0.6$), 
(4) Low ($0.2 \leq w < 0.4$), and 
(5) Very Low ($w < 0.2$). 
Figure~\ref{fig:flow_weighted_transp} shows the variation in traffic flow in the normalized, flow-weighted transportation network.


\subsection{Centrality Metrics}\label{sec:centrality_metrics}


\textcolor{black}{To identify critical nodes, we compute graph centrality metrics that rank nodes by structural importance. These rankings define targeted node-removal sequences to compare robustness under different notions of node criticality.}

\subsubsection{Degree Centrality: Identifying Network Hubs}
Degree centrality measures the number of direct connections a node possesses. For a node $i$ in an unweighted graph, degree centrality is the simplest and computationally most efficient centrality metric \citep{barabasi2016network}.

The degree of node $i$ is defined as:
\begin{equation}
k_i = \sum_{j=1}^{n} A_{ij}
\label{eq:degree_unweighted}
\end{equation}

where $A_{ij}$ is the adjacency matrix element (1 if an edge exists between nodes $i$ and $j$, 0 otherwise), and $n$ is the total number of nodes in the network. 

Degree centrality is defined as:
\begin{equation}
C_D(i) = \frac{k_i}{n-1}
\label{eq:degree_centrality}
\end{equation}

where $(n-1)$ is the maximum possible degree (a node connected to all others).

In road networks, degree centrality must identify major intersections and locations requiring traffic signal priority. In this case, edge weighting is required. For weighted networks (described in section~\ref{sec:flow_weight}), edges carry numerical values representing traffic flow, and degree centrality accounts for this flow using edge weights. The weighted degree, often termed \textit{node strength}, is defined as:
\begin{equation}
s_i = \sum_{j=1}^{n} fw_{ij}
\label{eq:weighted_degree}
\end{equation}

where $fw_{ij}$ is the normalized flow weight of the edge between nodes $i$ and $j$, defined in equation~\ref{eq:flow_weight}.

\begin{equation}
s_i^{\text{transport}} = \sum_{j} \text{fw}_{ij}
\end{equation}

Weighted degree centrality is:
\begin{equation}
C_D^{fw}(i) = \frac{\sum_{j=1}^{n} fw_{ij}}{\max_i \sum_{j=1}^{n} fw_{ij}}
\label{eq:weighted_degree_centrality}
\end{equation}

Removing high-degree nodes causes immediate local disconnection. However, the global impact of node removal depends on network topology factors such as in scale free networks \citep{Warner2019}, in which a few high-degree nodes support network connectivity and removing the top few nodes can fragment the network significantly.

\subsubsection{Closeness Centrality: Measuring Network Efficiency}\label{sec:closeness_centrality}

Closeness centrality measures how efficiently a node can reach all other nodes in the network. Removing high-closeness nodes increases average path length across the network. 
Nodes with high closeness are centrally positioned, with shorter average distances to other nodes. The shortest path distance from node $i$ to node $j$ is denoted $d(i,j)$ and represents the minimum number of edges required to traverse from $i$ to $j$ \citep{barabasi2016network}. 
For an unweighted network, this is computed using a breadth-first search.

Closeness centrality is defined as the inverse of the average distance:

\begin{equation}
C_C(i) = \frac{n-1}{\sum_{j=1}^{n} d(i,j)}
\label{eq:closeness_centrality}
\end{equation}

For weighted networks, shortest path distances must account for edge weights. 
The definition of distance depends on how weights are interpreted. 
When weights represent flow (higher weight = greater importance), distance is inversely proportional to weight:
\begin{equation}
d_{\text{fw}}(i,j) = \frac{1}{fw_{ij}}
\label{eq:distance_capacity}
\end{equation}

High-capacity edges are treated as ``short'' because they facilitate efficient flow. 
Closeness centrality for capacity-weighted networks is:
\begin{equation}
    C_C^{fw}(i) = \frac{n-1}{\sum_{j=1}^{n} d_{\text{fw}}(i,j)}
\label{eq:closeness_capacity}
\end{equation}

\subsubsection{Betweenness Centrality: Identifying Critical ``Bridges''}\label{sec:betweenness_centrality}

Betweenness centrality measures how frequently a node lies on shortest paths between other nodes. For example, an intersection with $C_B = 0.15$ has $15\%$ of the city's shortest paths passing through it. 
\textcolor{black}{Its removal forces rerouting of about $15\%$ of shortest-path traffic, indicating a critical bridge or bottleneck. In road networks, high-betweenness nodes often correspond to key intersections whose failure induces long detours, making them useful targets for bottleneck screening and emergency response planning.}

For an unweighted network, betweenness centrality is based on shortest path dependencies. The number of shortest paths from node $s$ to node $t$ through node $i$ is denoted $\sigma_{st}(i)$, while $\sigma_{st}$ denotes the total number of shortest paths from $s$ to $t$ \citep{barabasi2016network}.
Betweenness centrality of node $i$ is defined as:

\begin{equation}
C_B(i) = \sum_{s \neq i \neq t} \frac{\sigma_{st}(i)}{\sigma_{st}}
\label{eq:betweenness_centrality}
\end{equation}

where the sum is over all pairs of distinct nodes $(s, t)$ with $s \neq i \neq t$.

For weighted networks, betweenness centrality must account for the impact of weights on shortest path computation. The weighted shortest path distance is:
\begin{equation}
d_{fw}(s,t) = \min_{\text{path}} \sum_{(i,j) \in \text{path}} fw_{ij}
\end{equation}

Weighted betweenness centrality is defined using weighted shortest paths:
\begin{equation}
C_B^w(i) = \sum_{s \neq i \neq t} \frac{\sigma_{st}^w(i)}{\sigma_{st}^w}
\label{eq:betweenness_weighted}
\end{equation}

where $\sigma_{st}^w(i)$ is the number of weighted shortest paths from $s$ to $t$ passing through $i$, and $\sigma_{st}^w$ is the total number of weighted shortest paths from $s$ to $t$.

When weights represent traffic flow (higher weight = higher importance), the weight is inverted to define distance:
\begin{equation}
d_{\text{fw}}(i,j) = \frac{1}{fw_{ij}}
\label{eq:betweenness_flow_distance}
\end{equation}

High-flow edges are treated as ``short'', so shortest paths preferentially route through high-flow channels:
\begin{equation}
C_B^{\text{fw}}(i) = \sum_{s \neq i \neq t} \frac{\sigma_{st}^{\text{fw}}(i)}{\sigma_{st}^{\text{fw}}}
\end{equation}

\subsubsection{Eigenvector Centrality: Measuring Influence and Propagation}\label{sec:eigenvector}

\textcolor{black}{Eigenvector centrality measures node importance recursively: a node is influential if it connects to other influential nodes. In road networks, it highlights intersections tied to major hubs, even if they have modest degree, and is useful for studying influence propagation and diffusion \citep{barabasi2016network}.}

Eigenvector centrality is derived from the properties of the eigenvectors
of the network adjacency matrix. The adjacency matrix $\mathbf{A}$ is an $n \times n$ matrix where:
\begin{equation}
A_{ij} = \begin{cases} 1 & \text{if edge }(i,j)\text{ exists} \\ 0 & \text{otherwise} \end{cases}
\end{equation}

The eigenvector centrality $\mathbf{x}$ is defined as the principal eigenvector (eigenvector corresponding to the largest eigenvalue $\lambda_{\max}$) of the adjacency matrix:
\begin{equation}
\mathbf{A} \cdot \mathbf{x} = \lambda_{\max} \cdot \mathbf{x}
\label{eq:eigenvector_equation}
\end{equation}

For a node $i$, the eigenvector centrality is the $i$-th component of this eigenvector:
\begin{equation}
C_E(i) = x_i
\label{eq:eigenvector_centrality_simple}
\end{equation}

The eigenvector equation can be expressed recursively. 
The eigenvalue equation $\mathbf{A} \mathbf{x} = \lambda_{\max} \mathbf{x}$ implies:
\begin{equation}
x_i = \frac{1}{\lambda_{\max}} \sum_{j=1}^{n} A_{ij} \cdot x_j
\label{eq:eigenvector_recursive}
\end{equation}

This equation shows that the centrality of node $i$ is proportional to the sum of centralities of its neighbors. Nodes connected to high-centrality nodes receive high centrality themselves.

For weighted networks, the adjacency matrix is replaced with the weight matrix $\mathbf{W}$, where $W_{ij} = {fw}_{ij}$ is the edge weight:
\begin{equation}
\mathbf{W} \cdot \mathbf{x}^{fw} = \lambda_{\max} \cdot \mathbf{x}^{fw}
\label{eq:eigenvector_weighted}
\end{equation}

The weighted eigenvector centrality becomes:
\begin{equation}
C_E^{fw}(i) = \frac{1}{\lambda_{\max}} \sum_{j=1}^{n} fw_{ij} \cdot C_E^{fw}(j)
\label{eq:eigenvector_weighted_recursive}
\end{equation}

For flow-based weights, this captures connections to important high-flow neighboring intersections. Table~\ref{tab:centrality} summaries the comparison of the purpose of centrality measures under Unweighted and Weighted Graph.

\subsection{Resilience Metrics}\label{sec:reslience_metrics}
\textcolor{black}{Network resilience is evaluated using complementary structural and functional measures \citep{barabasi2016network}. Structural resilience is quantified by the giant connected component (GCC), i.e., the largest set of mutually reachable nodes remaining as nodes/edges are removed, representing the surviving connectivity backbone.}


\textcolor{black}{We define the State of Critical Functionality (SCF) as the fraction of original nodes remaining in the giant connected component after removing a fraction $f$ of nodes \citep{Bhatia_2015_PlosOne}:}
\begin{equation}\label{GCC}
\textcolor{black}{SCF=\frac{|GCC_f|}{|TT|}}
\end{equation}
\textcolor{black}{where $|GCC_f|$ is the GCC size after removals and $|TT|$ is the total node count in the undisrupted network. For flow robustness, SCF is similarly defined as the ratio of remaining network flow capacity to the original capacity, quantifying progressive degradation under cascading failures.}




\section{Resilience Analysis Results}\label{sec:results}

\subsection{Network Topology }\label{sec:topology}

\begin{figure}[H]
  \centering

  \begin{subfigure}[b]{0.49\textwidth}
    \centering
    \includegraphics[width=\textwidth, trim=0cm 0cm 0cm 0cm, clip]{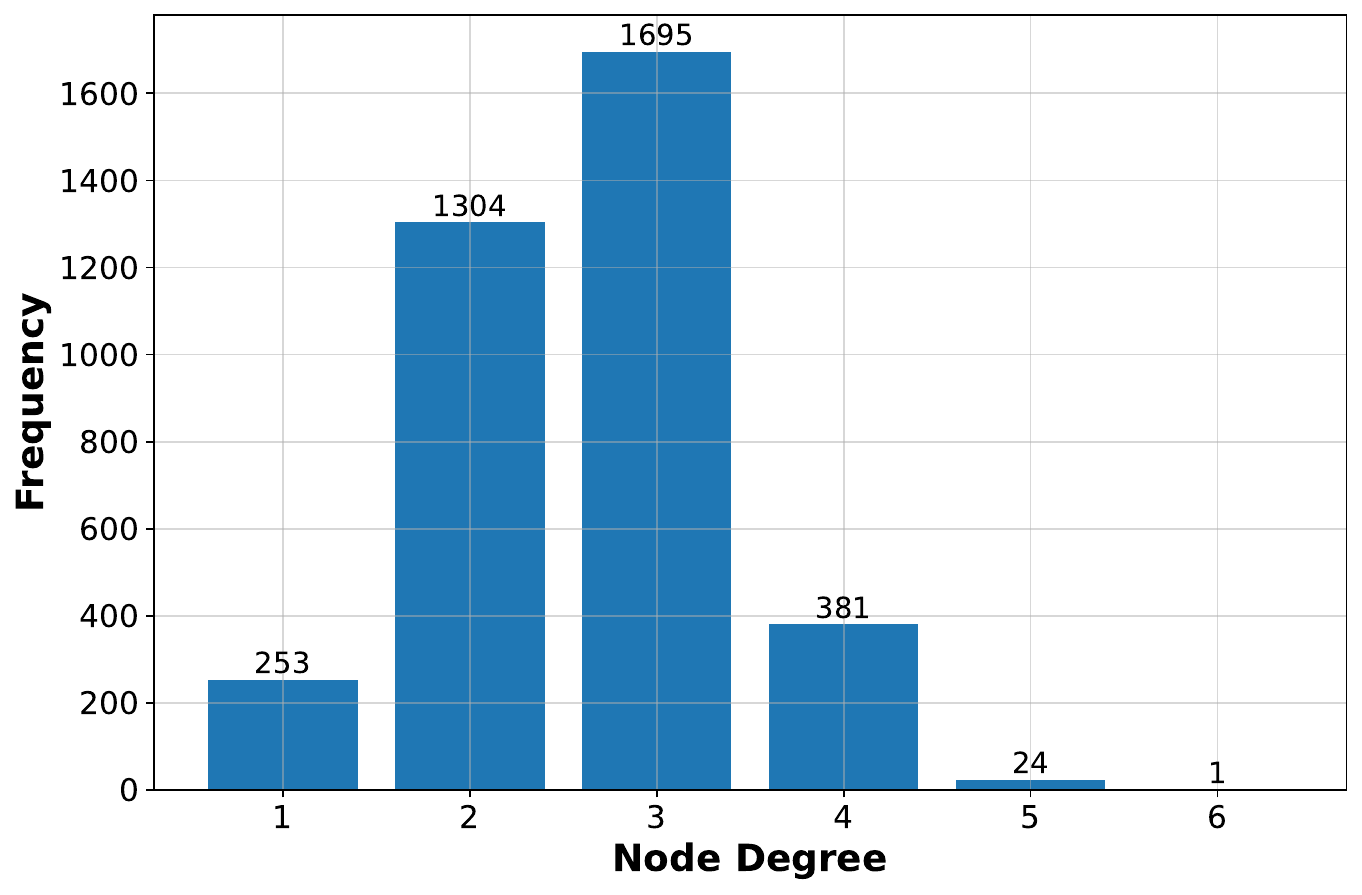}
    \caption{}
    \label{fig:BS_DegreeDist}
  \end{subfigure}
  \hfill
    \begin{subfigure}[b]{0.49\textwidth}
    \centering
    \includegraphics[width=\textwidth, trim=0cm 0cm 0cm 0cm, clip]{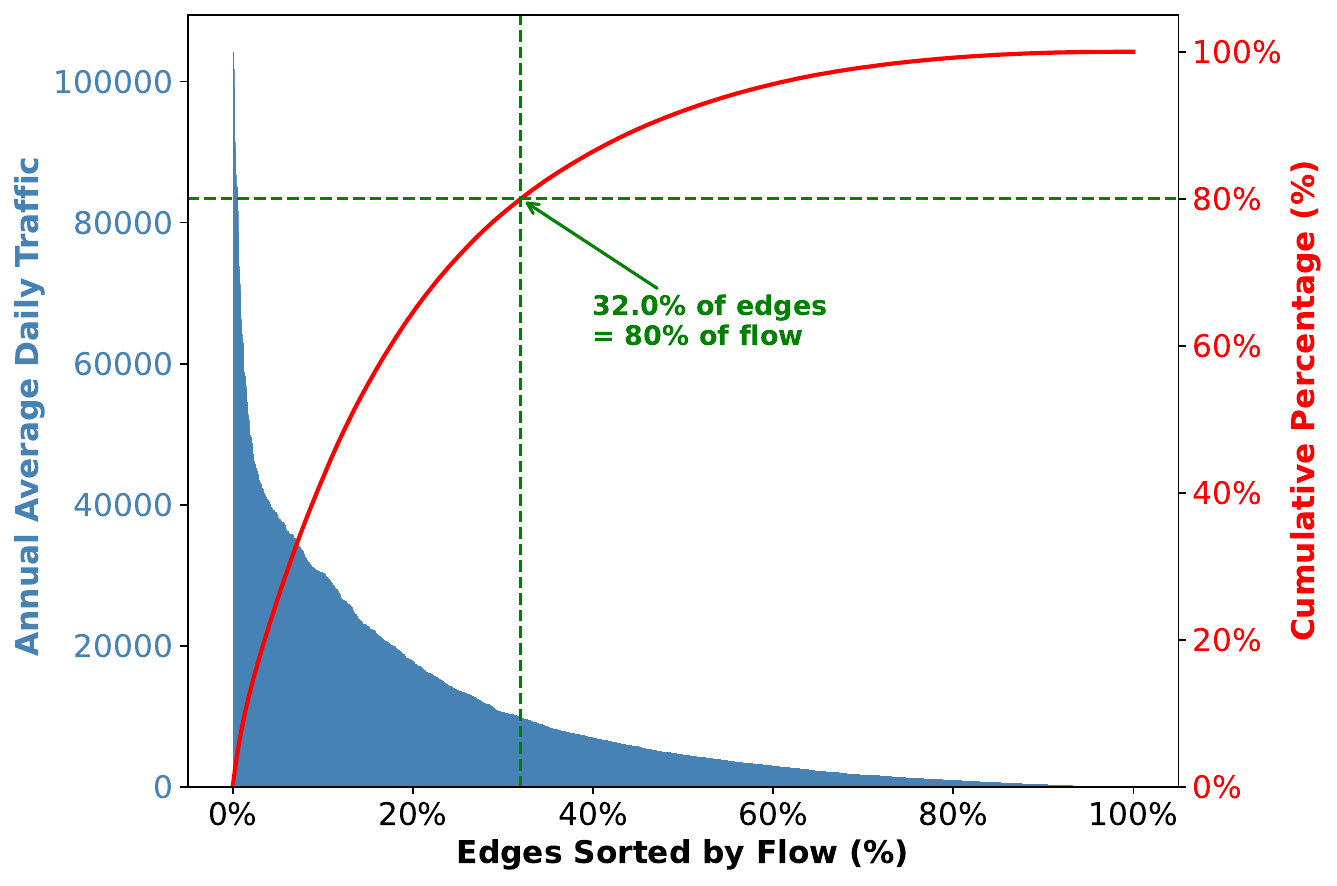} 
    \caption{}
    \label{fig:BS_ParetoFlowDist_b}
  \end{subfigure}

  \caption{(a) Degree distribution of the transportation network, illustrating heterogeneous connectivity. (b) Annual average daily traffic flow distribution, highlighting a Pareto distribution of traffic flow with a shape factor of 1.24.}
  \label{fig:degree_dist_pair}
\end{figure}

\textcolor{black}{Network topology provides a first-order view of structural vulnerability. The transportation network has a degree distribution (Fig.~\ref{fig:BS_DegreeDist}) peaked at $k=3$: 1,695 nodes (46.3\%) are T-intersections, 1,304 (35.6\%) have $k=2$, 381 (10.4\%) have $k=4$, 253 (6.9\%) are dead-ends ($k=1$), and only 25 (0.7\%) have $k=5$--6. The average degree is $\langle k\rangle=2.62$. The total flow across edges follows a Pareto distribution (see Figure~\ref{fig:BS_ParetoFlowDist_b}) where 32\% of nodes account for 80\% of networks' AADT}.

\subsection{Robustness Comparison: Weighted and Unweighted network}\label{sec:robustness_comparison}

\begin{figure}[H]
  \centering
  \begin{subfigure}[b]{0.49\textwidth}
    \includegraphics[width=\textwidth, trim=0cm 0cm 1cm 0cm, clip]{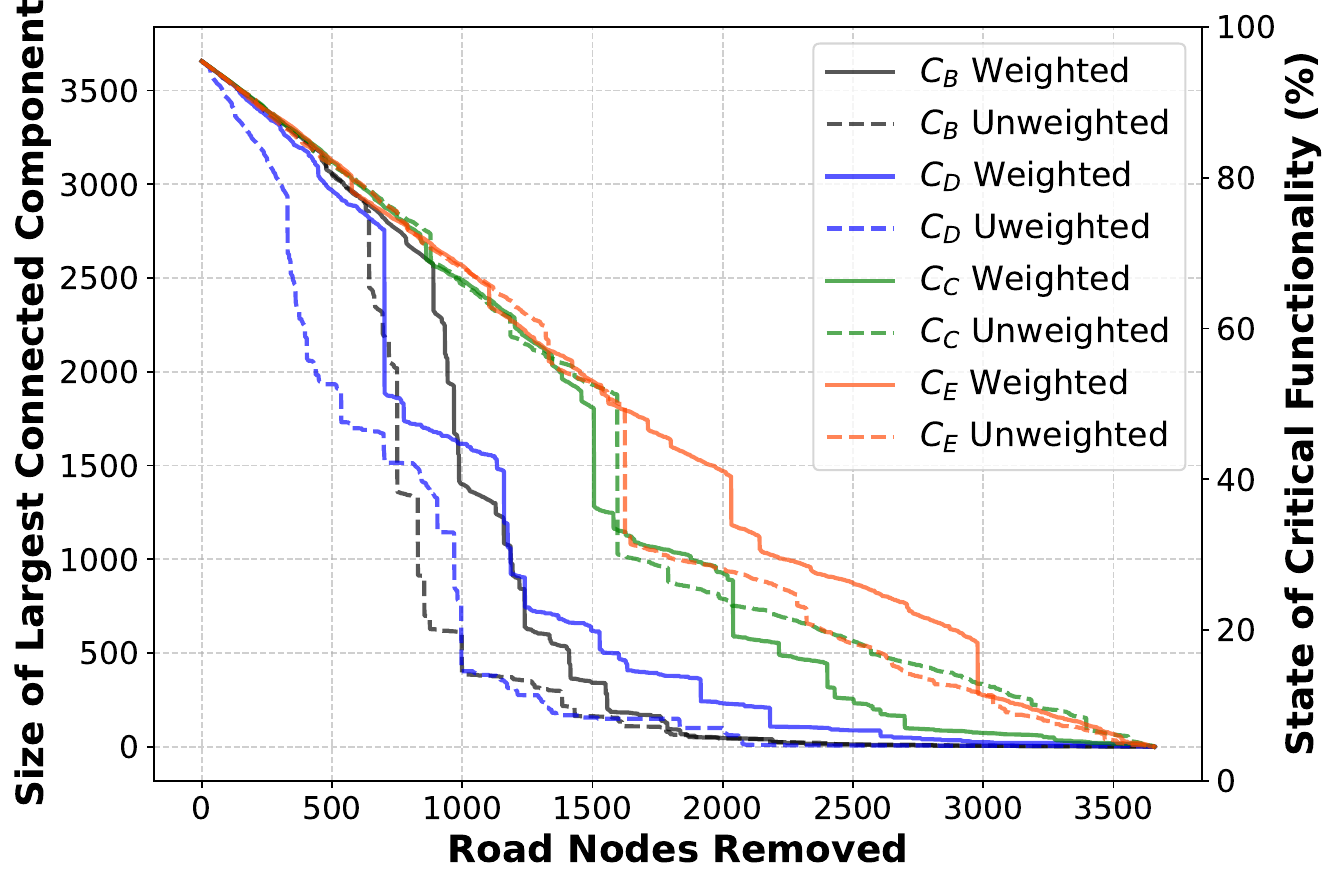} 
    \caption{}
    \label{fig:BS_Robustness_Initial_UnVsWeighted}
  \end{subfigure}
  \hfill
  \begin{subfigure}[b]{0.49\textwidth}
    \centering
    \includegraphics[width=\textwidth, trim=1cm 0cm 0cm 0cm, clip]{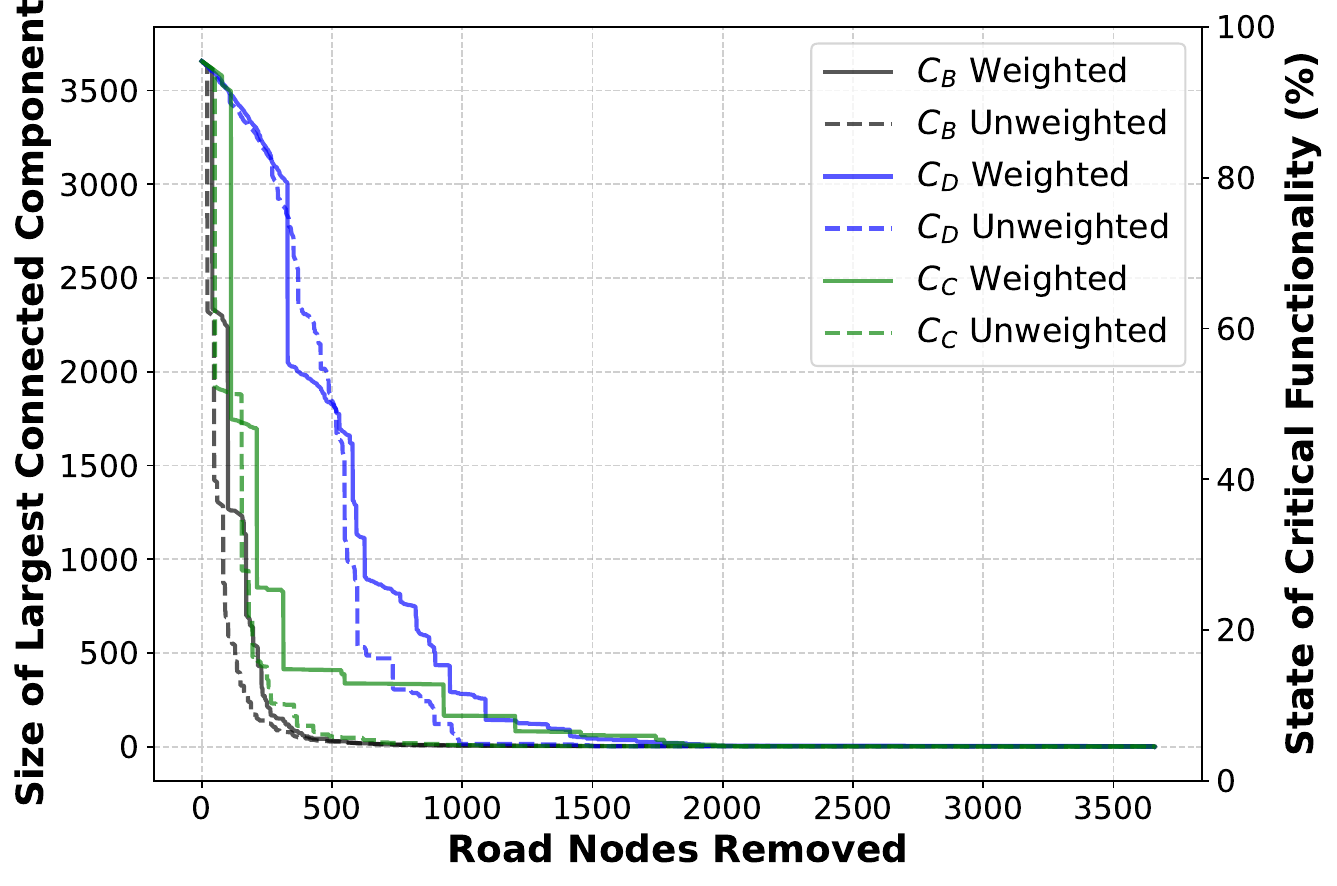} 
    \caption{}
    \label{fig:BS_Robustness_ReCal_UnVsWeighted}
  \end{subfigure}

  \begin{subfigure}[b]{0.49\textwidth}
    \includegraphics[width=\textwidth, trim=0cm 0cm 1cm 0cm, clip]{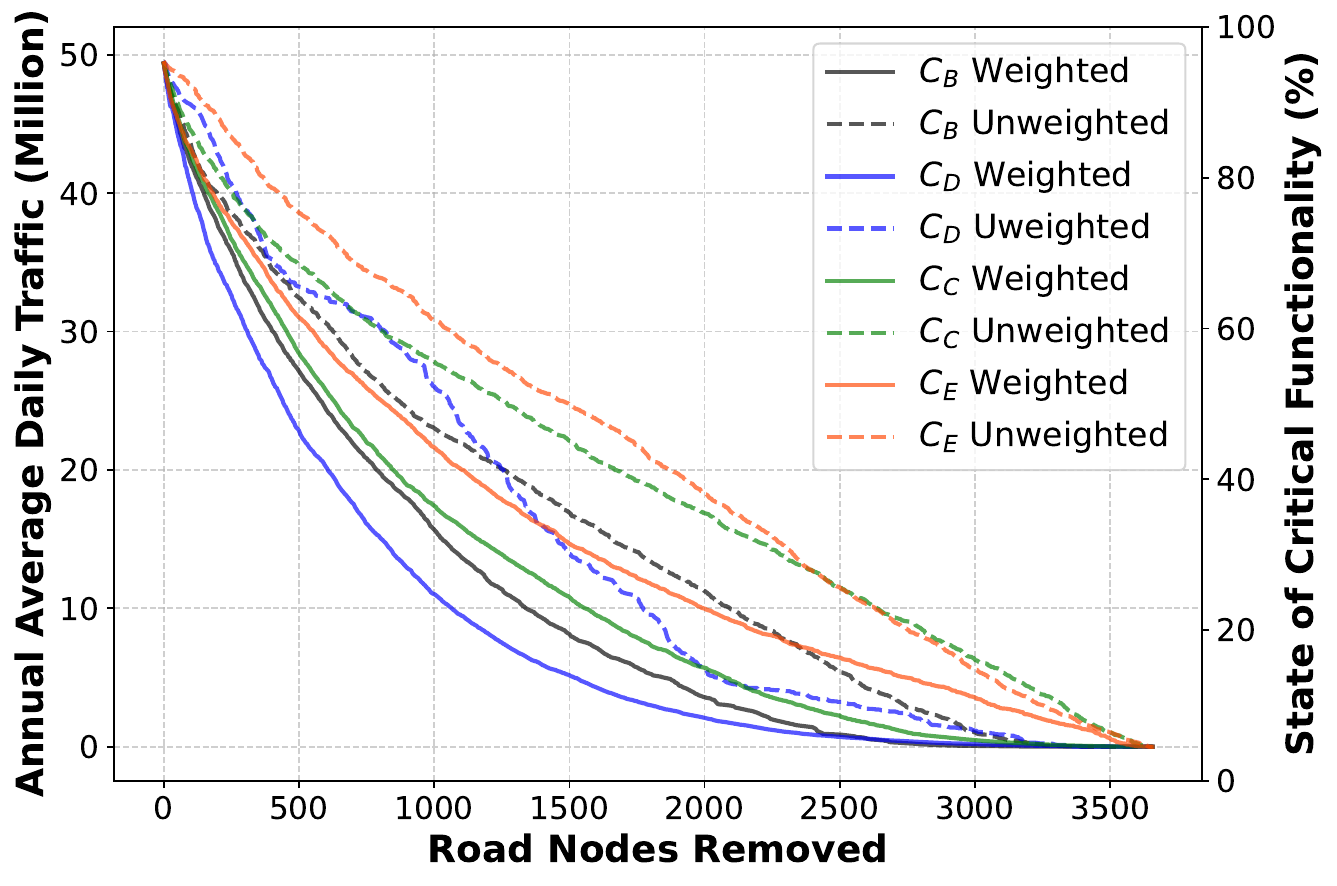} 
    \caption{}
    \label{fig:BS_Robustness_Initial_UnVsWeighted_Flow}
  \end{subfigure}
  \hfill
  \begin{subfigure}[b]{0.49\textwidth}
    \centering
    \includegraphics[width=\textwidth, trim=1cm 0cm 0cm 0cm, clip]{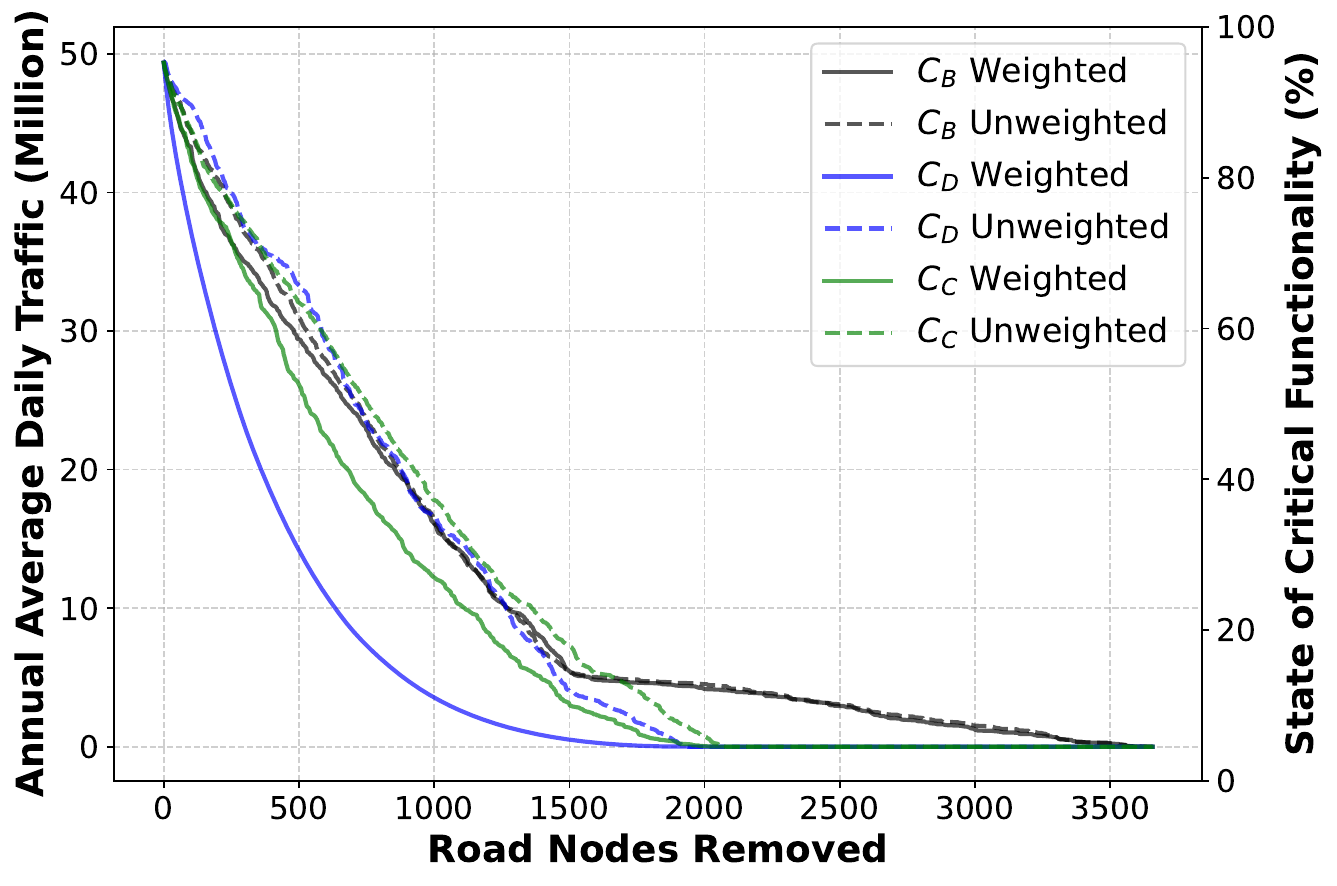} 
    \caption{}
    \label{fig:BS_Robustness_ReCal_UnVsWeighted_Flow}
  \end{subfigure}

  \caption{
  \textbf{Robustness of transportation network.} (a) Robustness analysis of unweighted vs weighted road transportation network using static centrality metrics. (b) Robustness analysis of unweighted vs weighted road transportation network using dynamic centrality metrics.
  (c) Impact on the annual average daily traffic count corresponding to the reduction of the GCC as shown in Figure~\ref{fig:BS_Robustness_Initial_UnVsWeighted}. (d) The drastic reduction in annual average daily traffic under dynamic robustness as shown in Figure~\ref{fig:BS_Robustness_ReCal_UnVsWeighted}.
  }
  \label{fig:Robustness_UnVsWeighted}
\end{figure}

Comparison of weighted and unweighted analyses (Figure~\ref{fig:Robustness_UnVsWeighted}) is essential to distinguish structurally central nodes from functionally critical ones and to assess how flow heterogeneity alters network response under disruption.


The robustness analysis of the GCC revealed that weighted transportation networks exhibit substantially greater resilience compared to their unweighted counterparts under both static and dynamic centrality-based disruption scenarios. 
In the initial or static centrality scenario (Figure~\ref{fig:BS_Robustness_Initial_UnVsWeighted}), weighted networks maintained larger GCC sizes across the entire range of node removals, with the performance gap becoming particularly pronounced after approximately 500--1000 nodes removal. 
The weighted network using betweenness centrality sustained a GCC size exceeding 1000 nodes and even after 1500 removals for other centrality metrics, whereas the corresponding unweighted network fragmented to below 500 nodes at the same removal threshold. 
This pattern persisted across all centrality metrics examined, indicating that the superiority of weighted networks is a general phenomenon rather than an artifact of specific targeting strategies \citep{Wang2008Universal, Bertagnolli2020Flow}. 
Under dynamic betweenness-based disruption scenarios (Figure~\ref{fig:BS_Robustness_ReCal_UnVsWeighted}), the weighted network retained connectivity for nearly twice as many node removals before complete collapse, suggesting that the functional redundancy encoded in edge weights becomes increasingly critical as the network structure evolves during sequential disruptions \citep{Sharmin2025Dynamic, Chen2021Assessing}. 
To provide additional context, Figure~\ref{fig:Robustness_All_strategies} presents robustness curves for unweighted and weighted transportation networks across all removal strategies. 
Consistent with the main results, dynamic centrality leads to more rapid fragmentation than static rankings in both cases. \textcolor{black}{Our dynamic resilience calculations assume edge-level travel demand remains unchanged after node/edge removals. In practice, travelers reroute, and demands shift to alternate links; incorporating rerouting and demand redistribution will be part of our future work. This effort exposes the importance of going beyond structural or unweighted network analysis, especially in transportation networks.}
Across all strategies, unweighted networks exhibit steeper fragmentation and greater sensitivity to targeted disruption, while weighted networks maintain larger GCC sizes and collapse more slowly, reinforcing the resilience advantages identified above.

Figures~\ref{fig:BS_Robustness_Initial_UnVsWeighted_Flow} and~\ref{fig:BS_Robustness_ReCal_UnVsWeighted_Flow} quantify the functional impact by tracking AADT loss across the two frameworks. 
The weighted network centrality metrics effectively identify functionally critical nodes whose removal precipitates rapid AADT decline (Figure~\ref{fig:BS_Robustness_Initial_UnVsWeighted_Flow}).
This occurs because weighted metrics prioritize nodes that carry disproportionate traffic flow (see Figure~\ref{fig:BS_ParetoFlowDist_b}) rather than merely those with high connectivity; hence, we see faster reduction in AADT.
In contrast, unweighted centrality metrics rank nodes based purely on topological position, identifying structurally important nodes that effectively fragment the graph-theoretic connectivity of the network (Figure~\ref{fig:BS_Robustness_Initial_UnVsWeighted}).

Under dynamic failure, the AADT reduced rapidly by identifying flow carrying critical nodes, and the total flow reached zero for both $C_D$ and $C_B$ trajectories within 2000 node removal, essentially making the remaining network non-functional (Figure~\ref{fig:BS_Robustness_ReCal_UnVsWeighted_Flow}).
These results underscore the inadequacy of topological metrics alone in assessing transportation resilience. Incorporating flow dynamics reveals substantially greater fragility and more realistic failure thresholds.

\section{Variation in Network Fragmentation and Flow Persistence}

Figures~\ref{fig:Robustness_UnVsWeighted}(a) and (b) reveal abrupt fragmentation of the GCC, particularly under dynamic failures. However, this topological fragmentation masks an important phenomenon: total network flow decreases much more gradually than the network fragmentation (Figure~\ref{fig:Robustness_UnVsWeighted}). This apparent paradox arises because network fragmentation does not immediately eliminate flow. Although connectivity is lost between primary components, the constituent fragments retain internal edge flow, and these distributed flows exhibit heavy-tailed distributions that persist across removal sequences~\citep{janicka2026heavytailsdynamicflow}. A dynamic visualization illustrating network topology evolution alongside corresponding changes in flow and GCC size under weighted betweenness centrality ($C^{fw}_B$) is available at \url{https://sharma-bharat.github.io/sims/WCB.gif}.

The above visualization also cogently explains the flow plateau visible in Figure~\ref{fig:BS_Robustness_ReCal_UnVsWeighted_Flow}. 
Flow exhibits a steep decline until approximately 1500 nodes are removed, after which it plateaus. 
At this threshold, the GCC is reduced to a single edge (size 2), rendering $C^{flow}_B$ values zero for all remaining nodes. Subsequent removals follow random edge deletion, no longer targeting high-flow nodes. 
In contrast, under weighted degree centrality ($C^{fw}_D$), total flow reaches zero near 1500 removals, as shown in \url{https://sharma-bharat.github.io/sims/WCD.gif}.

The distinction between $C_D$ (unweighted degree) and $C^{fw}_D$ (weighted degree, also termed ``strength centrality'') illustrates this difference clearly. 
$C_D$ ranks nodes by topological importance—the number of connected edges—while $C^{fw}_D$ ranks nodes by flow capacity, identifying those that channel the largest traffic volumes. 
Consequently, the failure trajectories diverge sharply: $C_D$ disruptions cause slower functional degradation despite rapid GCC fragmentation (\url{https://sharma-bharat.github.io/sims/UnWCD.gif}), whereas $C^{flow}_D$ disruptions produce steep reductions in network flow capacity, directly targeting the corridors sustaining mobility.


\subsection{Network-on-network impact: Transportation and Grid}\label{sec:network-on-network}

\begin{figure}[!ht]
  \centering

  \begin{subfigure}[b]{0.98\textwidth}
    \centering
    \includegraphics[width=\textwidth]{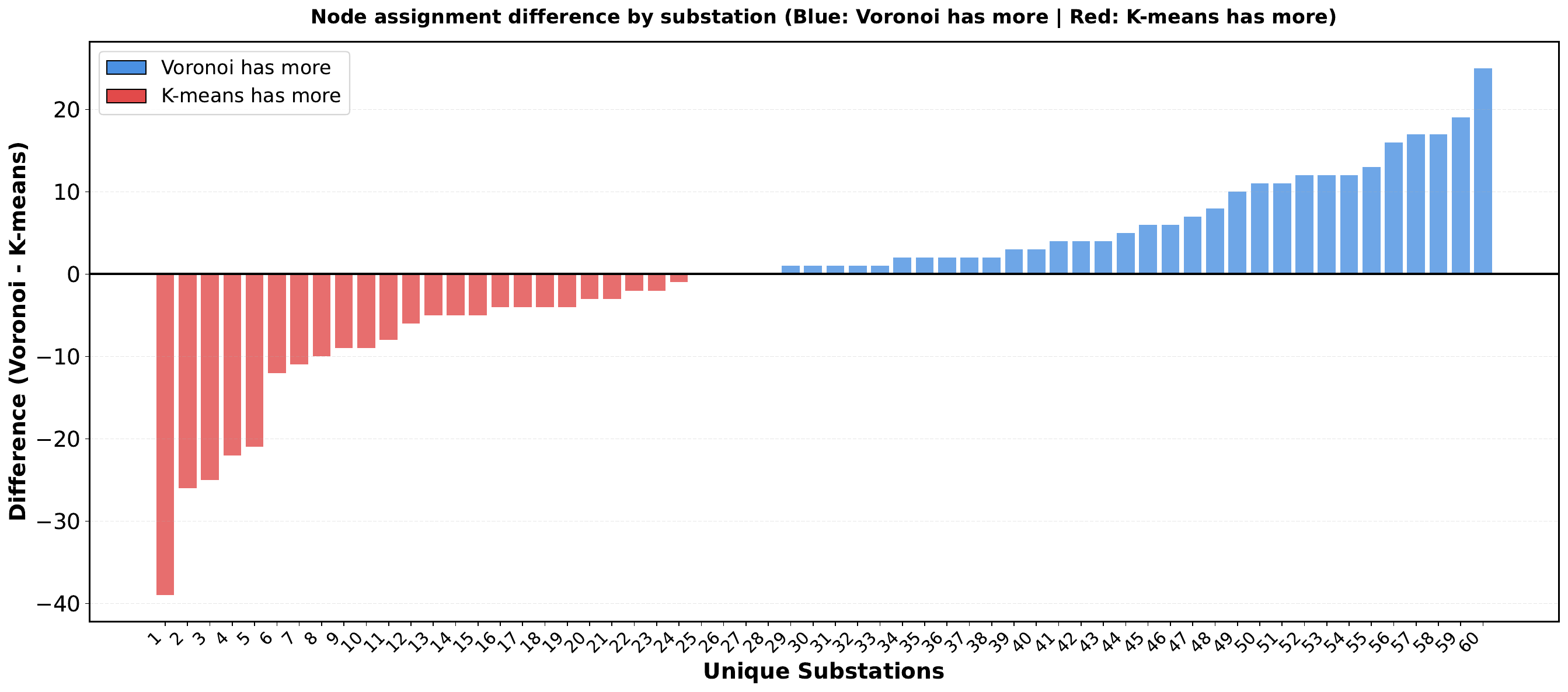}
    \caption{}
    \label{fig:voronoi_kmeans_difference}
  \end{subfigure}

  \vspace{0.6em}

  \begin{subfigure}[b]{0.49\textwidth}
    \centering
    \includegraphics[width=\textwidth, trim=0cm 0cm 1cm 0cm, clip]{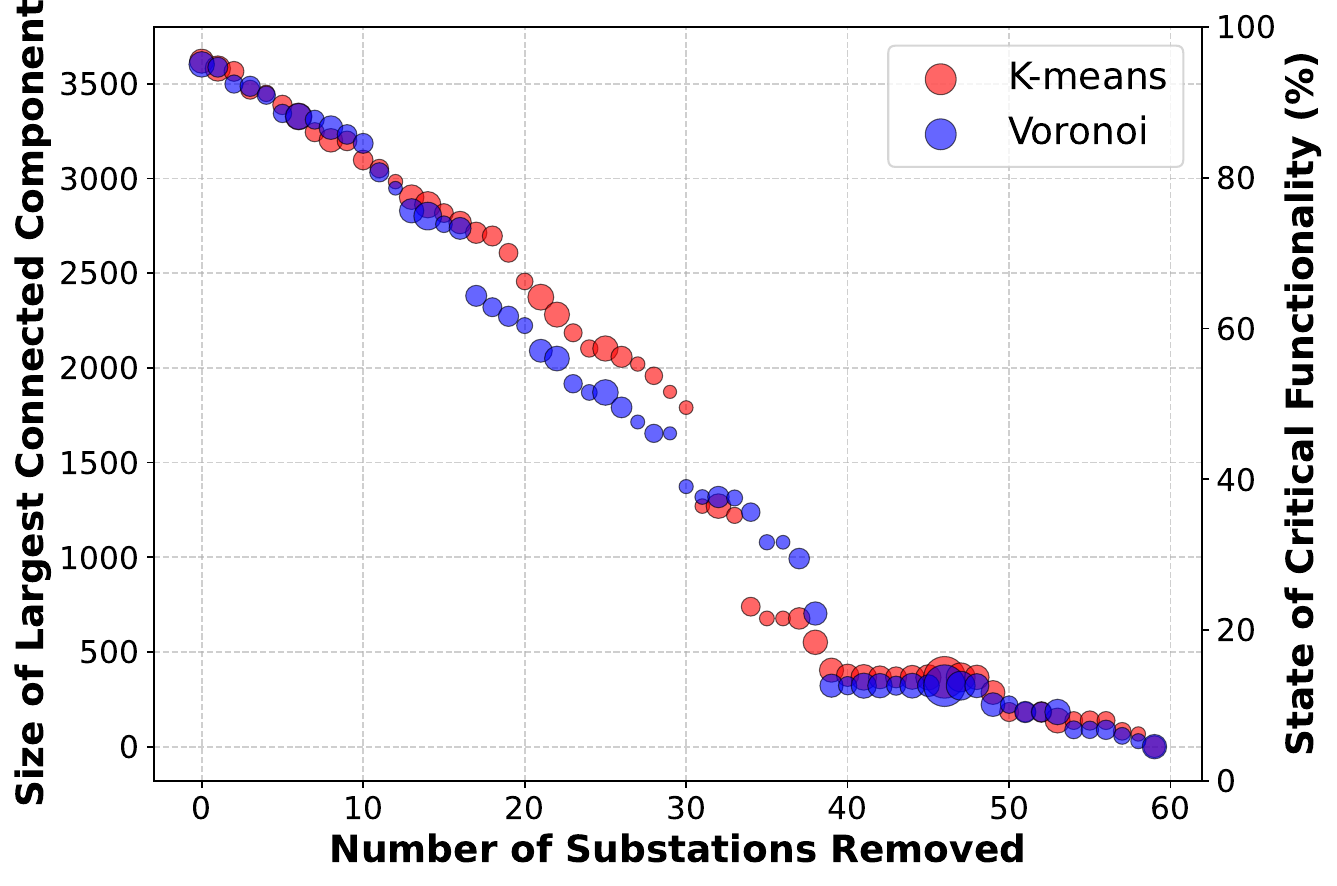}
    \caption{}
    \label{fig:BS_Random_Substation_Removal}
  \end{subfigure}
  \hfill
  \begin{subfigure}[b]{0.49\textwidth}
    \centering
    \includegraphics[width=\textwidth, trim=1cm 0cm 0cm 0cm, clip]{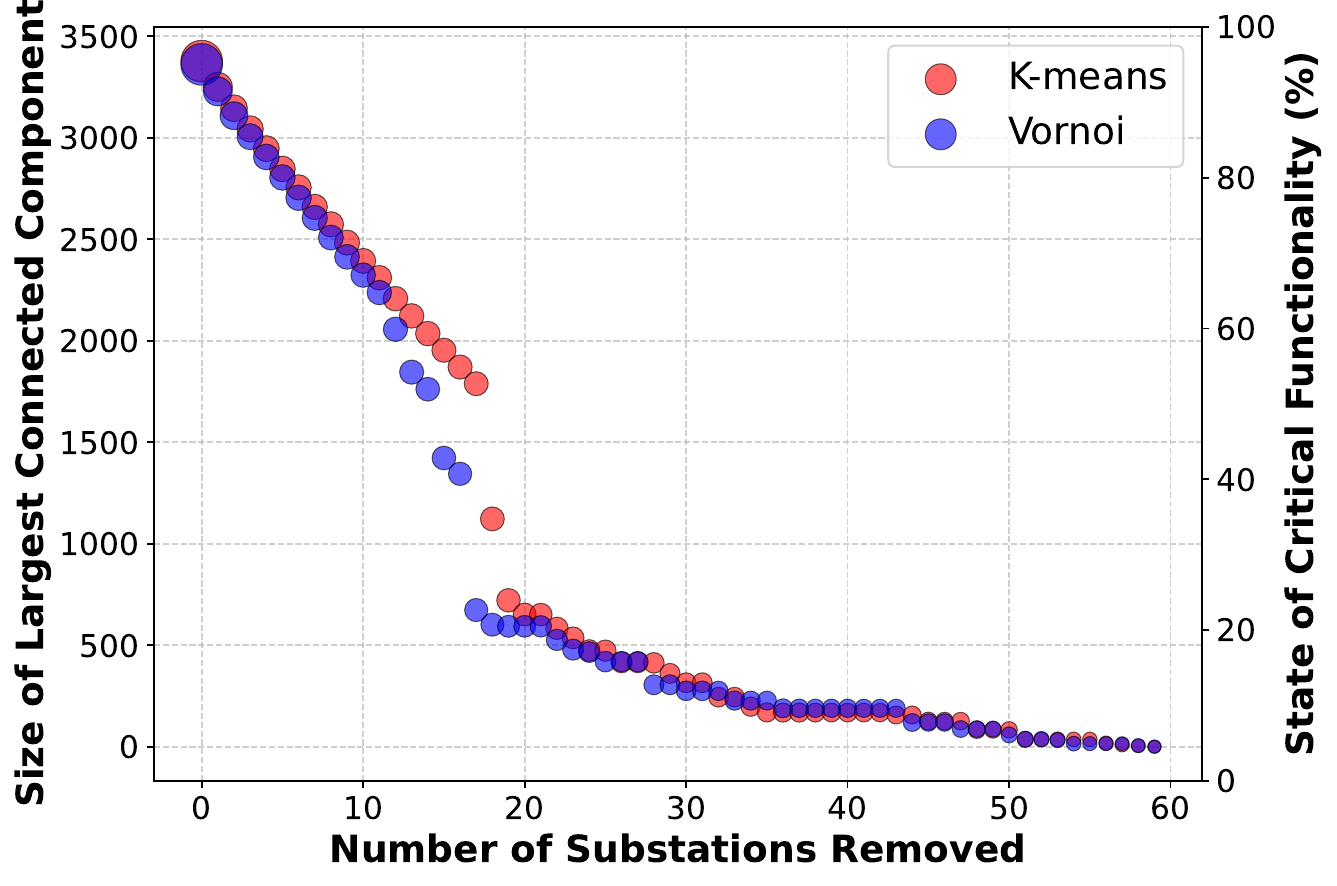}
    \caption{}
    \label{fig:BS_DescendingClusterSize_Substation_Removal}
  \end{subfigure}

  \caption{
  \textbf{Robustness of interconnected substation and transportation network.} (a) Node-assignment difference by substation between Voronoi and K-means, computed as $\Delta = N_{\mathrm{Voronoi}} - N_{\mathrm{K\text{-}means}}$. Impact of substation removal on the road network in (b) random order and (c) descending order of cluster size.
  }
  \label{fig:Robustness_NON}
\end{figure}

  
For coupling the transportation network with electric substations, we estimated each substation's impact extent using two spatial partitioning methods: K-means clustering and Voronoi tessellation. Transportation network nodes were assigned to a corresponding K-means cluster or Voronoi territory, implying that nodes within a territory are primarily served by the associated substation. Figure~\ref{fig:voronoi_kmeans_difference} compares these two assignments by plotting, for each unique substation, the signed difference in the number of assigned transportation nodes, $\Delta = N_{\mathrm{Voronoi}} - N_{\mathrm{K\text{-}means}}$. Negative (red) bars denote substations for which K-means assigns more nodes than Voronoi, whereas positive (blue) bars denote substations for which Voronoi assigns more nodes than K-means. Several substations exhibit disagreement, ranging from approximately $-39$ (K-means assigns substantially more nodes) to $+25$ (Voronoi assigns substantially more nodes). While most substations show differences near zero, a smaller subset displays large discrepancies. This comparison indicates that the choice of partitioning method can materially affect substation-level node counts. We next evaluated whether these assignment differences translate into meaningful changes in network robustness by conducting robustness analyses under both partitioning schemes.

Figures~\ref{fig:BS_Random_Substation_Removal} and~\ref{fig:BS_DescendingClusterSize_Substation_Removal} present the robustness analysis of the interconnected substation–transportation network under two substation failure strategies: 
(a) random removal, shown in Figure~\ref{fig:BS_Random_Substation_Removal}, and 
(b) targeted removal in descending order of cluster size, shown in Figure~\ref{fig:BS_DescendingClusterSize_Substation_Removal}.
In both cases, failure of a substation results in the removal of its associated cluster of road network nodes. 
The robustness analysis of the interconnected substation-transportation network reveals substantially different vulnerability profiles compared to the isolated road transportation network. 
Under random substation removal (Figure~\ref{fig:BS_Random_Substation_Removal}), the coupled network exhibited a relatively gradual decline in giant connected component size, maintaining approximately 2000 nodes after 20 substation failures and sustaining connectivity until approximately 50 substations were removed. 
The network degradation followed a near-linear pattern initially, suggesting that random failures in the electric infrastructure caused localized disruptions in dependent road clusters without triggering immediate systemic collapse. 
This resilience can be attributed to the distributed spatial configuration of substations serving geographically dispersed road network clusters, wherein the failure of a single substation affects only its associated cluster rather than creating network-wide cascading effects. 
By contrast, when substations were removed in descending order of cluster size (Figure~\ref{fig:BS_DescendingClusterSize_Substation_Removal}), the coupled network experienced dramatically accelerated fragmentation. This finding emphasizes the necessity of redundant pathways particularly in coupled urban networks.
\textcolor{black}{The giant connected component (GCC)} collapsed to $\sim$10\% of its original size after removing only 25--30 substations, indicating a vulnerability threshold three orders of magnitude more severe than random disruption. This brittleness under cluster-size-based removal reflects the hierarchical service architecture of electric--transport coupling: large clusters concentrate dependencies on a few high-capacity substations that serve densely connected road regions. This finding emphasizes the necessity of redundant pathways particularly in coupled urban networks.

\section{Discussion and Conclusion}\label{sec:discussion_conclusion}

The superior robustness of weighted networks stems from the fundamental difference in how topological versus functional importance is encoded and targeted during sequential node removal. 
In unweighted networks, centrality metrics rely exclusively on structural properties which treat all edges as equivalent \citep{Zhang2021Analysis}. 
The unweighted centrality approach systematically overestimates the criticality of structurally well-connected nodes that may carry relatively low traffic volumes, leading to premature targeting of nodes whose removal has limited impact on actual network flow capacity \citep{Wang2022Dynamic}. 
In contrast, weighted centrality measures incorporate traffic flow, capacity, or travel time into shortest path calculations, prioritizing nodes and edges that serve as high-capacity corridors in the functional flow network \citep{Bertagnolli2020Flow, Wang2008Universal}. 
These high-weight paths typically possess greater redundancy in the form of parallel medium-capacity routes, alternative connections, and distributed load absorption mechanisms that preserve connectivity even as critical hubs are removed \citep{Asztalos2011Optimal, Wang2008Universal}. 
When centrality is recalculated dynamically after each removal, the weighted network continues to identify functionally critical elements while the embedded flow distribution allows remaining nodes to reroute demand through secondary corridors, maintaining a viable connected component until a substantially larger fraction of the infrastructure has been compromised \citep{Sharmin2025Dynamic, Chen2021Assessing}. 
\textcolor{black}{This finding underscores that vulnerability assessment and resilience planning should prioritize flow-aware metrics over purely topological measures, since operational robustness depends on the network’s ability to absorb and redistribute traffic rather than structural connectivity alone \citep{Zhang2021Analysis}. Introducing cross-system dependencies exposes the limits of internal robustness \citep{buldyrev2010catastrophic}: when transportation is coupled to energy infrastructure, node failures depend not only on within-network structure and flow redistribution, but also on power availability. In this coupled setting, failures propagate through shared functional dependencies, producing vulnerability patterns that cannot be inferred from isolated network analyses \citep{liu2025resilience}.}

From an urban metabolism perspective, energy and mobility function as interacting anabolic subsystems that jointly sustain urban flows of people, materials, goods and services.
Random disruptions within the energy system can produce localized metabolic injuries, disabling specific mobility clusters while allowing the broader system to continue operating \citep{guidotti2016modeling}. 
This behavior reflects a modular metabolic organization within the city, in which spatially distributed service areas limit the spread of disruptions and allow unaffected regions to continue to support circulation and exchange \citep{lordan2014robustness}. 
In contrast, targeted disruptions affecting energy service points that support the largest mobility clusters expose a pronounced metabolic fragility  \citep{buldyrev2010catastrophic, gao2016universal}. 
These clusters correspond to zones of concentrated anabolic activity, where high mobility demand and energy consumption are tightly co-located \citep{kennedy2007changing}. 
The loss of such service nodes simultaneously constrains energy supply and mobility-related flows, abruptly degrading major circulation pathways. 
The resulting rapid fragmentation indicates that, despite internal redundancy within the transportation network, the coupled metabolic system relies on a few dominant service nodes with limited cross-system substitutability \citep{shamsi2025interdependency,gim2022institutional}. 
Taken together, these findings distinguish internal network robustness defined by connectivity within a single network from systemic metabolic resilience, which depends on the continuity of flows across interdependent urban infrastructures \citep{gao2016universal}. 
\textcolor{black}{While flow-aware weighting improves a transportation network’s ability to absorb disruptions, it does not ensure resilience at the urban metabolic scale \citep{tong2025resilience}. Urban resilience depends not only on within-network redundancy, but also on the structure, hierarchy, and spatial concentration of interdependencies between anabolic subsystems such as energy and mobility \citep{gim2022institutional}. Explicitly modeling both functional importance and cross-system coupling is essential to identify vulnerabilities, anticipate cascading failures, and support resilience planning that sustains and restores critical urban flows.}

\textcolor{black}{This study shows that urban infrastructure resilience assessments depend strongly on how functional importance and interdependencies are represented. Flow-aware, weighted analyses reveal substantial internal robustness in transportation networks due to redundant pathways. However, within a coupled energy--mobility system, this robustness is insufficient because hierarchical service dependencies can trigger rapid fragmentation under targeted disruptions.}
Interpreted through an urban metabolism lens, these findings underscore that resilience emerges from the coordinated performance of interacting anabolic subsystems rather than from the strength of individual networks in isolation. 
Urban metabolic stability therefore hinges not only on internal redundancy but also on the spatial organization, infrastructure service hierarchy, and coupling structure of critical infrastructures. 
Explicitly modeling these interactions is essential for accurately identifying vulnerability, anticipating cascading effects, and supporting resilience planning that safeguards the continuity of urban metabolic flows.

\section{Limitations and Future Work}\label{sec:limitation_future_work}
\textcolor{black}{This study is limited by using theoretical, centrality-driven removals rather than observed failures. Future work should incorporate empirically grounded hazards (e.g., flood footprints) with exact geographic locations to simulate spatially explicit disruptions.}
The transportation network assumes all nodes operate as signalized intersections; that is, they will be directly impacted due to substation failure. 
In practice, many nodes may be stop-controlled or uncontrolled, resulting in heterogeneous operational responses to loss of power. 
In the current scope of this paper, dynamic rerouting of traffic flow post-disruption is not recalculated due to the limitation of resources to run traffic simulation on a large urban network.
\textcolor{black}{Incorporating intersection control metadata (e.g., signalization status and timing plans) and explicitly modeling streetlight-outage effects would improve the realism of estimated mobility impacts \citep{saroj2021digitaltwin,xu2025automated}.}

\textcolor{black}{We approximate grid-transport interdependencies via spatial partitioning (k-means or Voronoi), treating substations as generator points and assigning each transportation node to its nearest substation (cluster/Voronoi territory). This captures spatial proximity and co-location but may not represent operational dependencies (e.g., feeder connectivity or the circuits powering intersections).}

\textcolor{black}{Future work should incorporate utility feeder-level connectivity to represent true grid-transport interdependencies and cascading failures. In addition, our analysis does not model traffic rerouting or resulting capacity/congestion on alternative links after node failures; integrating rerouting behavior with link-capacity constraints would better capture disruption-driven flow redistribution and mobility impacts.}

\clearpage

\section*{Data and Code Availability}

Data will be available upon request. 
The codes  will be made public after the acceptance of the manuscript.

\section*{Declaration}
The butterfly image in Figure 1 was generated by prompting ChatGPT.

\section*{Acknowledgment}
The authors thank the Transportation Planning Organization for raw traffic assignment and transportation network data. 

This manuscript has been authored by UT-Battelle LLC under contract DE-AC05-00OR22725 with the US Department of Energy (DOE). The US government retains and the publisher, by accepting the article for publication, acknowledges that the US government retains a nonexclusive, paid-up, irrevocable worldwide license to publish or reproduce the published form of this manuscript, or allow others to do so, for US government purposes. DOE will provide public access to these results of federally sponsored research in accordance with the DOE Public Access Plan (http://energy.gov/downloads/doe-public-access-plan).

\section*{CRediT authorship contribution statement}
Bharat Sharma: Writing – review \& editing, Writing – original draft, Visualization, Validation, Methodology, Investigation, Analysis, Conceptualization. 
Abhilasha~J.~Saroj: Writing – review \& editing, Conceptualization, Visualization, Analysis, Investigation. 
Evan Scherrer: Conceptualization \& Initial Results.
Melissa~R.~Allen-Dumas: Conceptualization, Investigation, Writing – review \& editing.

\clearpage

\appendix

\section{Supplementary Information}

\renewcommand{\thefigure}{A~\arabic{figure}}
\setcounter{figure}{0}
\renewcommand{\thetable}{A~\arabic{table}}
\setcounter{table}{0}

\begin{figure}[!ht]
  \centering
  \begin{subfigure}[b]{0.48\textwidth}
    \includegraphics[width=\textwidth, trim=0cm 0cm 0cm 1.5cm, clip]{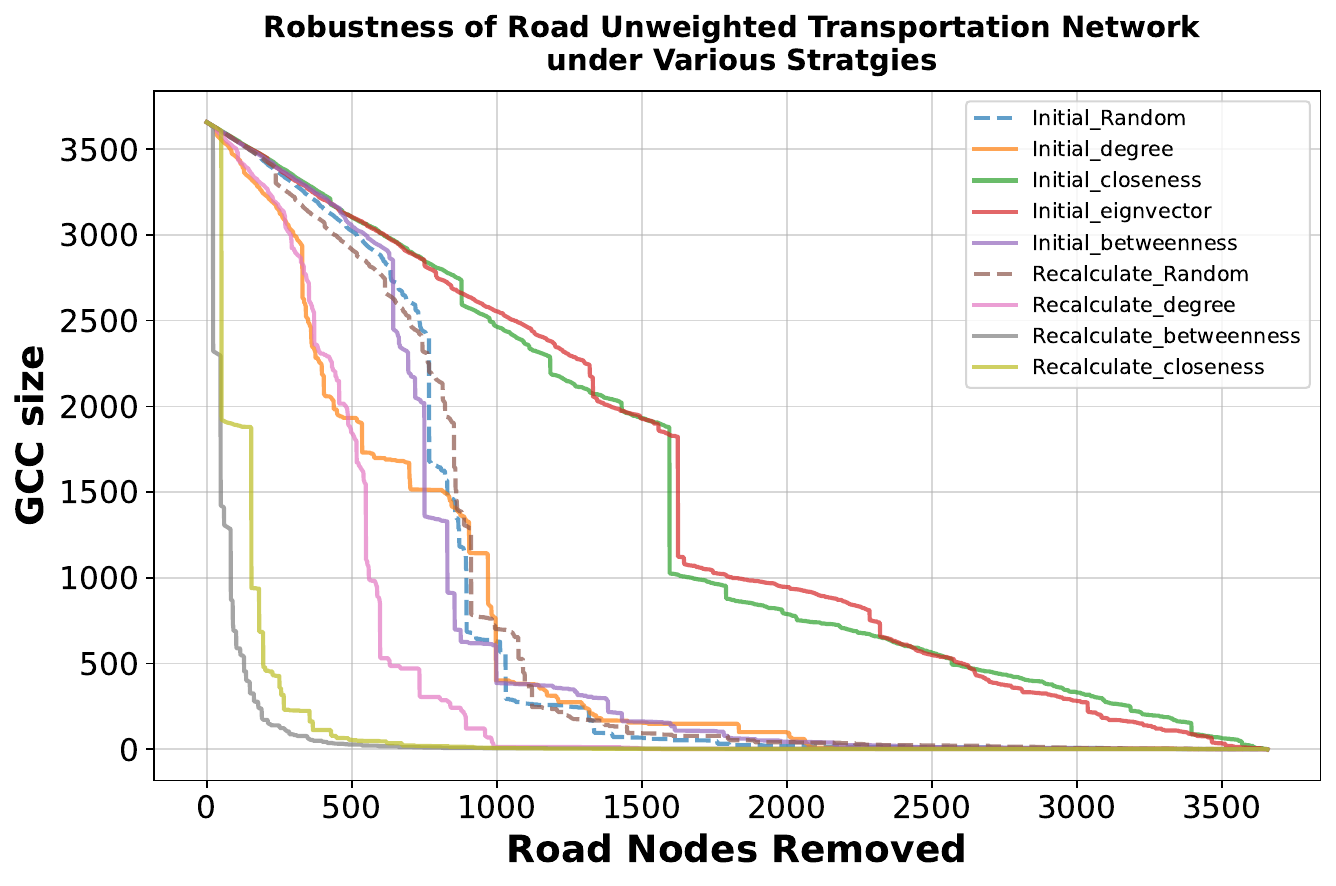} 
    \caption{}
    \label{fig:BS_Robustness_Unweighted}
  \end{subfigure}
  \hfill
  \begin{subfigure}[b]{0.48\textwidth}
    \centering
    \includegraphics[width=\textwidth, trim=0cm 0cm 0cm 1.5cm, clip]{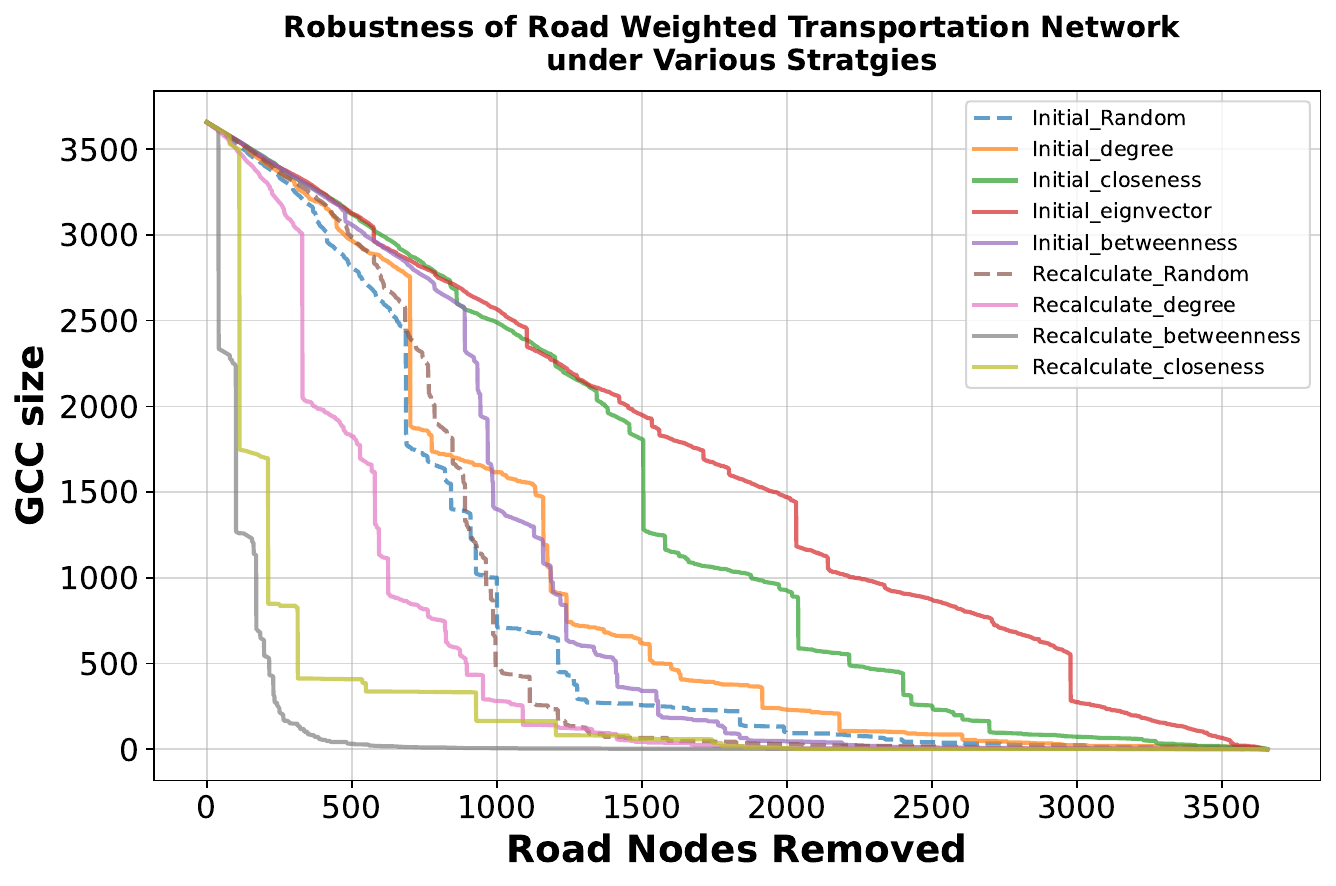} 
    \caption{}
    \label{fig:BS_Robustness_Weighted}
  \end{subfigure}
  
  \caption{
  \textbf{Robustness of interconnected substation and transportation network.} (a) Robustness of road unweighted transportation network under various strategies. Comparison of random, degree, closeness, eigenvector, and betweenness removal strategies for unweighted network analysis. Dynamic centrality metrics (recomputed after each removal) show substantially faster network collapse than static centrality metrics (fixed from baseline network. (b) Robustness of road weighted transportation network under various strategies. Same strategies as unweighted analysis but with weighted network analysis incorporating flow-based metrics. The unweighted network shows steeper fragmentation curves and greater sensitivity to targeted disruptions compared to the weighted case. 
  }
  \label{fig:Robustness_All_strategies}
\end{figure}

Figure~\ref{fig:Robustness_All_strategies} shows the robustness analysis of the GCC of unweighted and weighted network analysis under static and dynamic failure. We also show a baseline robustness to a random attack, which is not shown in the main manuscript due to a lack of reproducibility unless we define a particular seed value or a zone of randomness with multiple simulations. \textcolor{black}{Random node removal is not a traditional centrality metric but provides a key baseline for evaluating targeted disruption scenarios.}
In random removal, nodes are selected for removal with uniform probability, independent of their topological properties or functional importance \citep{barabasi2016network}. 
For a network with $n$ nodes, the probability of selecting any node $i$ for removal at step $t$ is:
\begin{equation}
P(\text{remove node } i \text{ at step } t) = \frac{1}{n - t + 1}
\label{eq:random_probability}
\end{equation}
where $t$ is the current removal step ($t = 1, 2, \ldots, n$).

The random centrality measure, could be represented as:
\begin{equation}
C_R(i) \sim \text{Uniform}(0, 1) \quad \forall i
\label{eq:random_centrality}
\end{equation}

All nodes have equal expected centrality under random removal. 
This contrasts with all other centrality metrics, which assign heterogeneous importance values.

\begin{table}[h]
\centering
\caption{Comprehensive Comparison of Centrality Metric}
\label{tab:centrality}
\resizebox{\columnwidth}{!}{%
\begin{tabular}{|c|c|c|c|}
\hline
\textbf{Centrality   Metric} & \textbf{Purpose (Unweighted)} & \textbf{Weight Type} & \textbf{Purpose (Weighted)} \\ \hline
Degree & Structural connectivity & Traffic\_Flow & Flow volume \& operational capacity \\ \hline
Betweenness & Shortest path bottlenecks & 1/Traffic\_Flow & Bottlenecks in efficient flow paths \\ \hline
Closeness & Accessibility \& reach & 1/Traffic\_Flow & Efficient accessibility via high-flow routes \\ \hline
Eigenvector & Connection to important neighbors & Traffic\_Flow & Hub importance in flow network \\ \hline
Randomness & Network topology Robustness Baseline & Traffic\_Flow & Flow distribution Baseline \\ \hline
\end{tabular}%
}
\end{table}

%


\begin{figure}[!ht]
  \centering
    \includegraphics[width=.9\textwidth, trim=0cm 0cm 0cm 0cm, clip]{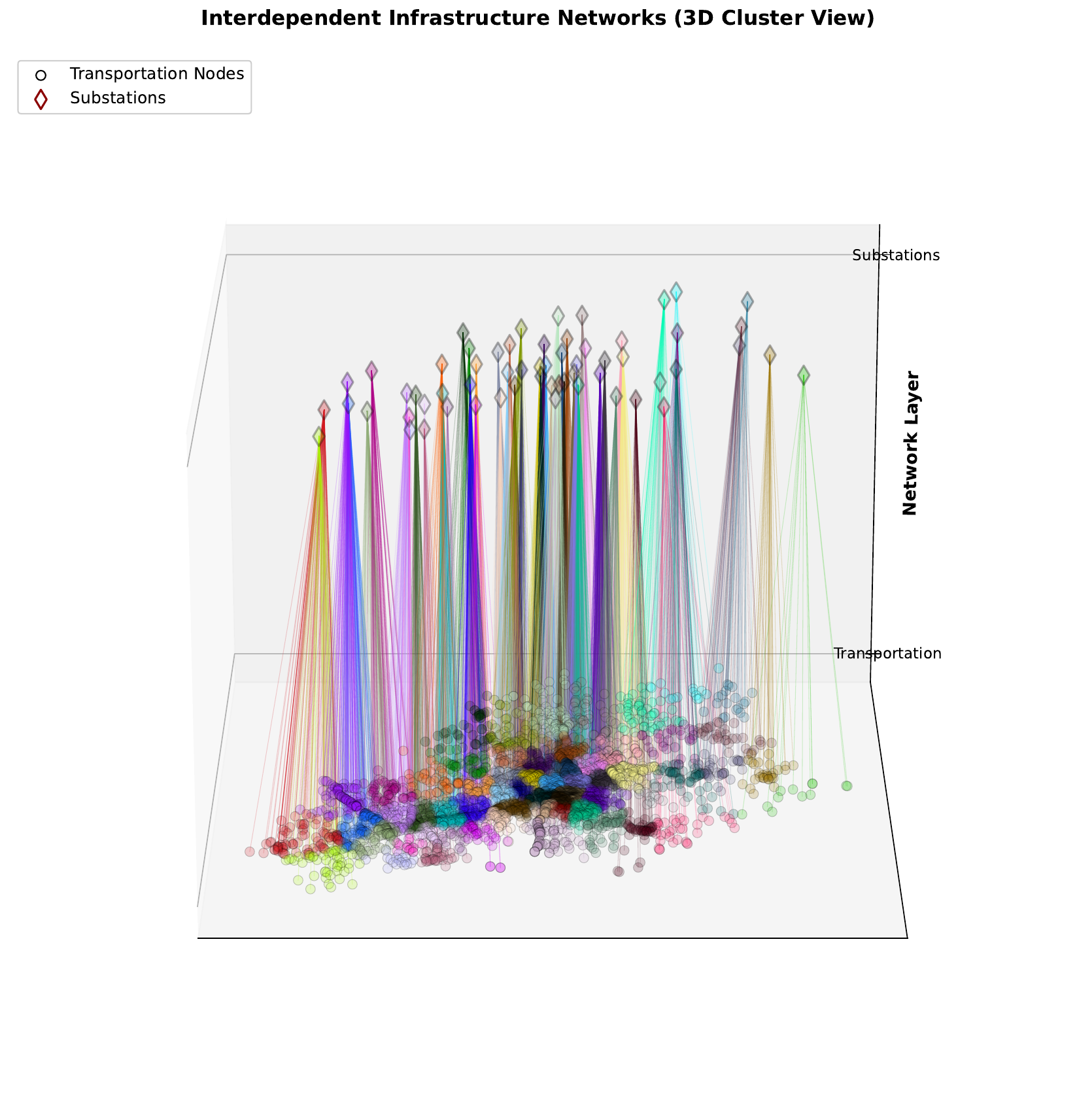} 
  
  \caption{
  \textbf{3D Schematic of Electric and Road Nodes}. 
  }
  \label{fig:ParetoDist_3d}
\end{figure}

\clearpage
\bibliographystyle{elsarticle-harv}
\bibliography{cas-refs.bib}

\end{document}